\documentclass[11pt,a4paper]{article}
\usepackage{jheppub}
\usepackage{mystyle}

\newcommand{\bO}{\boldsymbol{O}}

\newcommand{\bxi}{\boldsymbol{\xi}}
\newcommand{\eq}[1]{\begin{align}#1\end{align}}

\numberwithin{equation}{section}
\definecolor{LightGray}{rgb}{0.8, 0.8, 0.8}

\title{Conformal Blocks in 2d Carrollian/Galilean CFTs and Excited State Entanglement Entropy}
\author[a,c]{Peng-Xiang Hao}
\author[b,d]{Shunta Takahashi}

\affiliation[a]{
Center for Gravitational Physics and Quantum Information,
Yukawa Institute for Theoretical Physics, Kyoto University, Kyoto 606-8502, Japan}
\affiliation[b]{
Department of Physics, Kyoto University, Kyoto 606-8502, Japan
}
\affiliation[c]{Yau Mathematical Sciences Center, Tsinghua University, Haidian District, Beijing 100084, China}
\affiliation[d]{Perimeter Institute for Theoretical Physics, Waterloo, Ontario N2L 2Y5, Canada}
\date{}
\emailAdd{pxhao@yukawa.kyoto-u.ac.jp}
\emailAdd{shunta@gauge.scphys.kyoto-u.ac.jp}

\abstract{
    We advance the study of flat space holography by computing the entanglement entropy of highly excited states in two-dimensional Carrollian/Galilean Conformal Field Theories (C/G CFTs).
    Our approach is centered on a novel, physically intuitive derivation of the heavy-light conformal block in the large central charge limit, where the backreaction of heavy operators is absorbed by a C/G conformal coordinate transformation.
    Using this result and the replica trick, we find that the entanglement entropy of highly excited states assumes a thermal form, providing a concrete realization of the Eigenstate Thermalization Hypothesis (ETH).
    This field-theoretic result perfectly reproduces the holographic entanglement entropy computed via the swing surface proposal in three-dimensional Einstein gravity, for backgrounds corresponding to spinning particles and Flat Space Cosmological solutions.
    This agreement establishes a precise dictionary relating the weight $\Delta$ and charge $\xi$ of the boundary state to the mass $m$ and angular momentum $j$ of the dual spacetime, offering a powerful consistency check for the Flat/CCFT correspondence.
}


\begin{document}
    \begin{flushright}
    KUNS-3076
    \end{flushright}

    \maketitle

    \flushbottom

    \clearpage

    \section{Introduction}
        In pursuit of genuine quantum gravity, the emergence of spacetime is the most fundamental yet enigmatic conundrum, a solution to which remains an aspiration of humanity.
        RT formula \cite{Ryu:2006bv} in well-established AdS/CFT serves as a cornerstone that allows us to probe the bulk geometry of quantum gravity from quantum information theoretic quantity, namely the entanglement entropy, in boundary QFT without gravity.
        However, whether the narrative extends to more realistic model like flat holography or dS/CFT remains elusive.
        Towards the full picture of bulk emergence in flat holography, this paper aims to delineate the generic framework to compute entanglement entropy for broader class of asymptotically flat spacetimes from the boundary Carrollian/Galilean CFT.
        Specifically, we calculate it in the presence of highly excited states through conformal blocks analysis.

        Before discussing further details, let us first outline the recent developments in flat holography.
        The Flat/Carrollian CFT correspondence \cite{Barnich:2010eb, Bagchi:2010zz, Fareghbal:2013ifa,Bagchi:2025vri} is the holographic principle that relates the quantum gravity on asymptotically flat space with the Carrollian CFT (CCFT), also known as BMS field theory field theory, on the codimension one asymptotic null boundary with time direction.
        Despite several unique challenges including the peculiar Carrollian behavior \cite{bacry1968possible, bergshoeff2014dynamics, Duval:2014uoa} and the non-conserved gravitational charges \cite{Trautman:1958zdi, Wald:1999wa} due to the null boundary, Flat/CCFT has been the focus of intensive research in the past decade.
        As a holography, the Bekenstein-Hawking entropy of flat cosmological horizons matches with the Cardy formula in CCFTs \cite{Barnich:2012xq, Bagchi:2012xr}.
        The partition functions are discussed in \cite{Barnich:2012rz, barnich2015one}, and the one-loop determinant corresponds to the CCFT vacuum character.
        The bulk reconstruction is discussed \cite{Chen:2023naw, Hao:2025btl}, and the top-down construction from AdS/BCFT can be found in Ref.\,\cite{Hao:2025ocu}.
        The CCFT entanglement entropy is calculated by twist operator or Rindler method \cite{Bagchi:2014iea} and it can be compared to the swing surface proposal \cite{Jiang:2017ecm}, which generalizes the RT formula \cite{Ryu:2006bv} in AdS/CFT.
        Besides, CCFT itself is the target of interest as a special kind of quantum field, with the developments on the correlation functions \cite{Bagchi:2009ca}, bootstrap method \cite{Bagchi:2016geg, Bagchi:2017cpu, Chen:2020vvn, Chen:2022jhx}, the detailed scalar/fermion models \cite{Hao:2021urq, Chen:2021xkw, Yu:2022bcp, Hao:2022xhq, Banerjee:2022ocj, deBoer:2023fnj, Hao:2025hfa} among others.

        We note that there is another approach, celestial holography \cite{Pasterski:2016qvg, Pasterski:2017kqt}, where CFT exists on a codimensions \textit{two} boundary without time direction.
        Since gravitational $S$-matrix elements expressed in a boost eigenstate basis resemble conformal correlation functions, it excels at analyzing gravitational scattering problem.
        Its relation to Flat/CCFT has been gradually coming to light \cite{Donnay:2022aba, Donnay:2022wvx, Bagchi:2022emh, Bagchi:2023fbj}.
        However, we restrict our attention to Flat/CCFT and hereby present our new results on conformal block and entanglement entropy as a contribution to the established body of work.

        \subsubsection*{\colorbox{LightGray}{Main results}}
            \noindent\textbf{\underline{(i) Carrollian/Galilean conformal block}} \\
            We present the systematic and analytical calculation of conformal blocks for both light and heavy-light operator configurations in 2D Carrollian/Galilean CFTs at large central charge.
            A key finding is that, in the large central charge and heavy weight and charge limit,
            \begin{equation}
                c_M\rightarrow\infty,\ \ \Delta,\xi,\Delta_r,\xi_r\sim O(1),\ \ \frac{H}{c_M},\frac{\Xi}{c_M},\frac{c_L}{c_M}\sim O(1),
            \end{equation}
            the full C/G blocks reduce to global blocks through a novel C/G conformal transformation, with the explicit relation given by
            \begin{equation}
                \lim_{c_M\to\infty}g(x,y)=(1-w)^{\Delta(1-\frac{1}{\alpha})}\alpha^{2\Delta-\Delta_r}e^{\xi\frac{z(\alpha-1)}{(w-1)\alpha}}g_{\text{global}}(w,z).
            \end{equation}
            This transformation geometrically induces a rescaling and tilting of the cylinder, characterized by the parameters
            \begin{align}
                \alpha:=\sqrt{1-\frac{24\Xi}{c_M}},\quad Q:=\frac{(\alpha^2-1)c_L+24H}{2\alpha^2c_M}.
            \end{align}

            \noindent\textbf{\underline{(ii) Entanglement entropy in the presence of heavy states and holography}} \\
            Leveraging these results and the assumption of vacuum block dominance, we compute the entanglement entropy for states excited by a heavy operator.
            The final result for a single interval on the cylinder is
            \begin{equation}
                S_{A}=\frac{c_{L}}{6}\log\left(\frac{2}{\alpha\epsilon}\sin\frac{\alpha l_{\phi}}{2}\right)+\frac{c_{M}}{6}\left(Q+\frac{\alpha(l_{u}-Ql_{\phi})}{2}\cot\frac{\alpha l_{\phi}}{2}\right).
            \end{equation}
            We note that this result is identical to the vacuum entanglement entropy on a non-trivially identified cylinder.
            Furthermore, when the boost charge exceeds a specific threshold ($\Xi > c_M/24$), the parameter $\alpha$ becomes purely imaginary and the entropy assumes a thermal form, signaling the Eigenstate Thermalization Hypothesis (ETH) in these theories.

            Finally, we establish a precise holographic match.
            Our C/G CFT derived entanglement entropy for heavy states perfectly agrees with the holographic entanglement entropy computed via the swing surface proposal in 3D flat space, for either conical defect or Flat Space Cosmology (FSC) solutions (see eqs.\, \eqref{eq: hee for spining particle}, \eqref{eq: hee for fsc}).
            This agreement establishes a concrete dictionary that links the weight $\Delta$ and charge $\xi$ of the excited state to the angular momentum $j$ and mass $m$ of the dual spacetime.

        \subsection*{\colorbox{LightGray}{Relation to previous literature}}
            Our discussion is predicated on the pre-existing result in CFT$_2$ outlined in Section \ref{sec: ee at large central charge in cft}.
            The C/G conformal blocks are first obtained by Hijano in Ref.\,\cite{Hijano:2018nhq} via monodromy method, where they are determined by solving a second order differential equation that arises from the existence of null vector in the representation of Carrollian/Galilean conformal algebra \eqref{eq: BMSalg}.
            For better clarity, our contribution is systematically deriving the blocks by more physically intuitive way of Ref.\,\cite{Fitzpatrick:2015zha} in CFT$_2$.
            For simplicity, we consider the case where the propagating operator is a singlet under $M_0$.

            We then compute the block in certain large central charge limit to deduce the entanglement entropy in C/G CFT in the spirit of Refs.\ \cite{Hartman:2013mia, Asplund:2014coa}, where the conformal block in holographic CFT$_2$ in the large $c$ limit is computed.
            In $2$d Galilean/Carrolian CFT, the \textit{vacuum} entanglement entropy for single interval is already derived in Ref.\,\cite{Bagchi:2014iea}.
            Later, the RT formula \cite{Ryu:2006bv} in AdS/CFT is extended to flat holography version known as \textit{swing surface} proposal \cite{Jiang:2017ecm, Apolo:2020bld}, and the resulting holographic calculation for $3$d Minkowski spacetime agrees with the result of Ref.\,\cite{Bagchi:2014iea}.
            Our result serves as a further confirmation of swing surface proposal by showing the agreement of boundary heavy excited states entropy with the bulk holographic entropy for excited geometries, including FSC and conical defect solutions.

        \subsection*{\colorbox{LightGray}{Plan of the paper}}
\begin{figure}[h]
                \centering
                \begin{tikzpicture}[
                  node distance=20mm and 30mm,
                  box/.style={draw, rounded corners=2pt, very thick, align=center,
                              minimum width=45mm, minimum height=10mm}
                ]
                    \draw (-6.5,3) rectangle (-1.5,1.5);
                    \draw (6.5,3) rectangle (1.5,1.5);
                    \draw[fill=black!12] (-2.25,0.5) rectangle (2.25,-1);
                    \draw[fill=black!12] (-2.25,-2) rectangle (2.25,-3.5);
                    \draw (3.25,-2.5) rectangle (6.25,-4);

                    \node[above] at (-4, 2.25) {Section \ref{sec: ee at large central charge in cft}};
                    \node[below,scale=0.9] at (-4, 2.25) {Review of EE in AdS$_3$/CFT$_2$};
                    \node[above] at (4, 2.25) {Section \ref{sec: review of 2d ccft}};
                    \node[below,scale=0.9] at (4, 2.25) {Reveiw of C/G CFT};
                    \node[above] at (0, -0.25) {Section \ref{sec: bms block}};
                    \node[below,scale=0.9] at (0, -0.25) {C/G conformal block};
                    \node[above] at (0, -2.75) {Section \ref{sec: bms entanglement entropy}};
                    \node[below,scale=0.9] at (0, -2.75) {C/G entanglement entropy};
                    \node[above] at (4.75,-3.25) {Appendix \ref{append: swing surface}};
                    \node[below,scale=0.9] at (4.75,-3.25) {Swing surface};
                    \draw[->, thick] (-4, 1.5) -- (-2,-2);
                    \draw[->, thick] (2.25, 1.5) -- (0.5,0.5);
                    \draw[->, thick] (0, -1) -- (0,-2);
                    \draw[->, thick] (3.25, -3.25) -- (2.25,-2.75);
                \end{tikzpicture}
                \caption{Logical flow of this paper}
                \label{fig: paper flow}
        \end{figure}
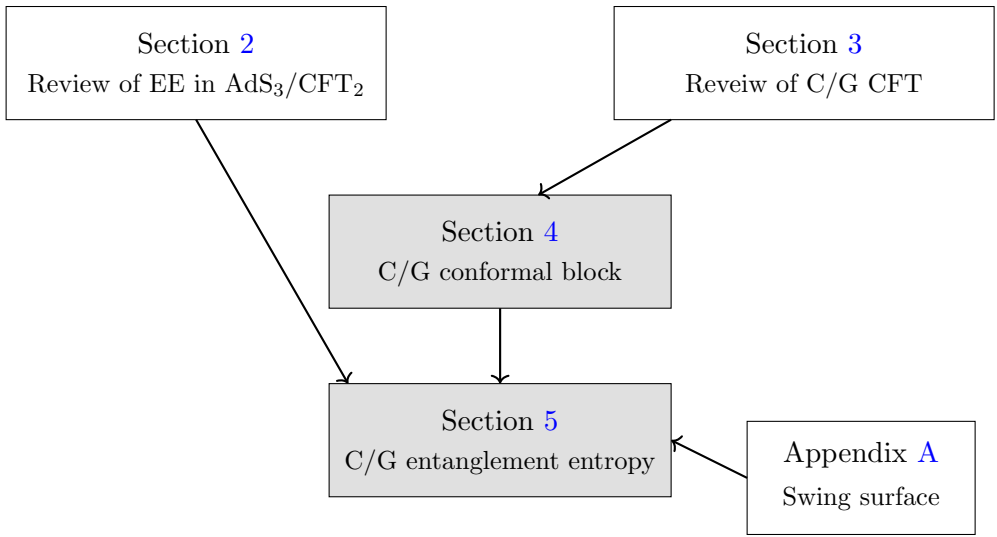

            This paper is organized as follows (See Fig.\,\ref{fig: paper flow} for the logical flow).
            In Section \ref{sec: ee at large central charge in cft}, we review entanglement entropy in 2D CFTs with large central charge and its holographic dual in AdS$_3$.
            In Section \ref{sec: review of 2d ccft},  we introduce the fundamentals of two-dimensional Carrollian/Galilean conformal field theories (CFTs), with a focus on the highest weight representation.
            Key topics include symmetries, stress tensors, correlation functions, and particularly the block expansion of four-point functions.

            Section \ref{sec: bms block} presents our main results.
            We derive conformal blocks for light and heavy-light operators in various large central charge limits, and demonstrate how the full blocks reduce to global blocks under specific C/G conformal transformations.
            Section \ref{sec: bms entanglement entropy} applies these blocks to compute excited state entanglement entropy and matches it with holographic results via the swing surface proposal, and establish a dictionary between C/G CFT heavy excitation and bulk spacetime parameters.

            Appendix \ref{append: swing surface} provides brief overview of the swing surface proposal for holographic entanglement entropy in asymptotically flat spacetimes, focusing on its geometric construction and relation to modular flow.

    \section{Review of entanglement entropy at large-$c$ in AdS$_3$/CFT$_2$} \label{sec: ee at large central charge in cft}
        In order to compute the entanglement entropy in $2$d Carrollian/Galilean CFT at ``large central charge'', we first recap the well-trodden case of CFT$_2$ in the large-$c$ limit and how it agrees with the semi-classical holographic entanglement entropy in the AdS$_3$ bulk.
        Inspired by the strategy in CFT$_2$, the following sections (Section \ref{sec: bms block} \& \ref{sec: bms entanglement entropy}) provide our new results in $2$d Carrollian/Galilean CFT.

        Returning to the review of CFT$_2$, the calculation simplifies dramatically in the limit of large central charge $c \gg 1$ and does not depend on the details of the spectrum (operator content and fusion rules) of CFT$_2$.
        This connection relies on the replica trick \cite{Calabrese:2009qy} and the universal behavior of conformal blocks \cite{Hartman:2013mia, Asplund:2014coa}.

        \subsection{Entanglement entropy from the replica trick at large $c$}
            The entanglement entropy $S_A:=-\Tr_A\rho_A\log\rho_A$ of a subsystem $A$ is calculated via the replica trick in CFT$_2$ \cite{Calabrese:2009qy}.
            To this end, one first computes the $n$-th R\'{e}nyi entropies
            \begin{align}
                S_A^{(n)} := -\frac{1}{n-1} \log \Tr_A\rho_A^n.
            \end{align}
            for integer $n \ge 2$, and then take the limit $n \to 1$ by viewing $n$ as a continuous variable
            \begin{equation}
                S_A = \lim_{n\to 1} S_A^{(n)}.
            \end{equation}
            For a single interval of length $L>0$ with its end points at $u,v\in\mathbb{C}$, one needs to prepare $n$ copies of the original CFT$_2$, glue them together on the intervals, and compute its partition function $Z_n$, so that $\rho_A^n=\frac{Z_n}{Z_1^n}$ is obtained.
            In CFT$_2$, however, $\Tr\rho_A^n$ for a single interval is equivalent to the two-point function of local twist operators $\mathcal{T}_n$ at the endpoints of the interval
            \begin{align}
                \operatorname{Tr}\rho_A^n=\langle\mathcal{T}_n(u)\mathcal{T}_n(v)\rangle,
            \end{align}
            where the conformal weight is $\Delta_n = \frac{c}{24}(n-\frac{1}{n})$.
            It yields
            \begin{equation}
                \Tr(\rho_A^n) \propto \left(\frac{L}{\epsilon}\right)^{-2\Delta_n} = \left(\frac{L}{\epsilon}\right)^{-\frac{c}{12}(n-1/n)},
            \end{equation}
            with $\epsilon$ a UV cutoff.
            The R\'{e}nyi entropy is
            \begin{equation}
                S_A^{(n)} = \frac{1}{1-n} \left(-\frac{c}{12}\left(n-\frac{1}{n}\right)\right)\log\left(\frac{L}{\epsilon}\right) = \frac{c}{12}\left(1+\frac{1}{n}\right)\log\left(\frac{L}{\epsilon}\right),
            \end{equation}
            and taking the limit $n \to 1$, one lands on the Calabrese-Cardy formula \cite{Calabrese:2009qy} for the vacuum entanglement entropy
            \begin{equation}
                S_A = \lim_{n\to 1} S_A^{(n)} = \frac{c}{6}\log\left(\frac{L}{\epsilon}\right).
            \end{equation}
            Note that the above formula captures only the holomorphic sector, and the full non-chiral result is $S_A = \frac{c}{3}\log\left(\frac{L}{\epsilon}\right).$

        \subsection{Excited state entropy from the heavy-light blocks and thermalization}
            We next consider the entanglement entropy in an excited state created by the insertion of a heavy local operator $O_H$, whose conformal dimension scales with the central charge, $h_H=O(c)$.
            The calculation of $\Tr\rho_A^n$ now requires computing a \textit{heavy-light type} four-point function of the form $\langle O_H(\infty) \mathcal{T}_n(u) \mathcal{T}_n(v) O_H(0)\rangle$.

            At large $c$, the heavy operators significantly back-react on the geometry.
            The implication of Ref.\,\cite{Asplund:2014coa} is that the heavy-light conformal block can be computed via a conformal transformation from the plane (the vacuum background) to the back-reacted geometry created by the heavy state $\displaystyle|H\rangle:=\lim_{z\to 0}O_H(z)\ket{0}$.
            The transformation is achieved by
            \begin{align}
                w(z) = z^\alpha,\quad\alpha = \sqrt{1 - \frac{24h_H}{c}}, \label{eq: mapping to conical deficit}
            \end{align}
            which maps the plane to a cone with a deficit angle $2\pi(1-\alpha)$.
            In the limit $1\ll h_L\ll c$, $h_H/c=O(1)$ the vacuum block is approximated by
            \begin{align}
                \mathcal{V}_0(u,v)\approx(\alpha)^{2h_L}v^{-h_L(1-\alpha)}\Big(\frac{1-v}{1-v^{\alpha}}\Big)^{2h_L}.
            \end{align}
            The replica calculation yields the entanglement entropy for an interval of length $L$
            \begin{equation}
                S_A =-\lim_{n\to\infty}\frac{1}{n-1}\bra{H}\mathcal{T}_n(u) \mathcal{T}_n(v)\ket{H}= \frac{c}{3} \log \left( \frac{\beta_{\text{eff}}}{\pi \epsilon} \sinh\left(\frac{\pi L}{\beta_{\text{eff}}}\right) \right),
            \end{equation}
            which is the standard formula for EE in a thermal CFT at temperature $T=1/\beta_{\text{eff}}$ with
            \begin{equation}
                \beta_{\text{eff}} = \frac{2\pi R_{\text{AdS}}}{\sqrt{\frac{24h_H}{c} - 1}}.  \label{eq: cft temperature}
            \end{equation}
            For high-energy states above the black hole threshold $h_H > c/24$, the parameter $\alpha$ becomes purely imaginary $\alpha = i\sqrt{\frac{24h_H}{c} - 1}$, and the mapping \eqref{eq: mapping to conical deficit} implies a periodic identification in the Euclidean time direction on the cylinder since
            \begin{align}
                z^{i\sqrt{\frac{24h_H}{c} - 1}}=\exp\bigg(\underbrace{i\sqrt{\frac{24h_H}{c} - 1}\log|z|}_{\text{time direction on cylinder}}-\sqrt{\frac{24h_H}{c} - 1}\arg z\bigg).
            \end{align}
            This periodicity provides the signature for a realization of the eigenstate thermalization hypothesis (ETH) in holographic CFTs, since the high-energy expectation value is indistinguishable from the thermal state with the effective inverse temperature \eqref{eq: cft temperature}.

        \subsection{Holographic match: the BTZ black hole}
            The holographic dual of a (spinless) heavy primary state $|H\rangle$ is a microstate of a (non-rotating) BTZ black hole
            \begin{align}
                g_{\text{BTZ}}=-\Big(\frac{r^2}{R_{\text{AdS}}^2}-8GM\Big)dt^2+\frac{1}{\frac{r^2}{R_{\text{AdS}}^2}-8GM}dr^2+r^2d\phi^2.  \label{eq: btz metric}
            \end{align}
            The mass of the BTZ black hole is directly related to the energy of the state
            \begin{align}
                M = \frac{2}{R_{\text{AdS}}}\Big(h_H - \frac{c}{24}\Big).  \label{eq: btz mass confomal weight}
            \end{align}
            To verify the CFT result, we apply the RT formula in the BTZ geometry. The task is to compute the length of a spatial geodesic anchored at the endpoints of the interval on the asymptotic boundary of the BTZ spacetime.

            This geodesic calculation from the BTZ metric \eqref{eq: btz metric} yields a length
            \begin{equation}
                A(\gamma_A) = 2 R_{\text{AdS}} \log \left( \frac{\beta_{\text{BTZ}}}{\pi \epsilon} \sinh\left(\frac{\pi L}{\beta_{\text{BTZ}}}\right) \right),
            \end{equation}
            where $\beta_{\text{BTZ}}=\frac{\pi R_{\text{AdS}}}{\sqrt{2GM}}$ is the inverse Hawking temperature of the black hole.
            Eqs.\,\eqref{eq: cft temperature}, \eqref{eq: btz mass confomal weight} reveal that $\beta_{\text{BTZ}} = R_{\text{AdS}}\beta_{\text{eff}}$, so applying the RT formula with the Brown-Henneaux relation $\frac{3R_{\text{AdS}}}{2G}$ again produces a perfect match with the CFT result for the heavy state.

    \section{2d Carrollian/Galilean CFTs in a nutshell} \label{sec: review of 2d ccft}
        Before embarking on the study of $2$d Carrollian/Galilean conformal block in analogy with CFT$_2$ result stated in the previous section, let us encapsulate the fundamental aspects of $2$d Carrollian/Galilean conformal field theories which will be used in the later discussions.

        The 2d Carrollian/Galilean (C/G) conformal field theories are quantum field theories that remain invariant under the following transformations,
        \eq{\widetilde{x}= f(x), \quad \widetilde{y}= f^{\prime}(x)y + g(x).\label{eq: BMSplane}}
        The generators implementing transformations on the fields are denoted as $ L_m $ and $ M_n $.
        These generators form the two-dimensional Carrollian/Galilean conformal algebra, also known as the BMS$_3$ algebra,
        \begin{align}
            \begin{aligned}
                \left[L_{n},\,L_{m}\right] & = (n-m)L_{m+n}+\frac{c_L}{12} n(n^2-1)\delta_{m+n,0}, \\
                \left[L_{n},\,M_{m}\right] & = (n-m)M_{m+n}+\frac{c_M}{12}  n(n^2-1)\delta_{m+n,0}, \\
                \left[M_{n},\,M_{m}\right] & = 0,
            \end{aligned}
            \label{eq: BMSalg}
        \end{align}
        where $c_L$ and $c_M$ are central charges.
        Consider the following infinitesimal translation
        \begin{align}
            \sigma\rightarrow \sigma + \varepsilon,\quad \tau\rightarrow\tau+\tilde{\varepsilon},
        \end{align}
        where $\varepsilon$ and $\tilde{\varepsilon}$ are constants.
        The Noether theorem allows us to derive two conserved currents,
        \begin{align}\label{cur}
            2\pi \pmb{j}_{\varepsilon} &= - \varepsilon T   d\sigma  + \varepsilon M d\tau,\\
            2\pi \pmb{j}_{\tilde{\varepsilon}} &= -\tilde{\varepsilon}M  d\sigma,
        \end{align}
        where $T$ and $M$ play the role of the stress tensors.
        Then the infinitely many charges on the plane can be written as
        \eq{
            L_n&=\frac{1}{2\pi i}\oint \Big( x^{n+1} T+  (n+1)x^{n}y M \Big)dx,\\
            M_n&= \frac{1}{2\pi i}\oint x^{n+1} Mdx, \label{modes1}
        }
        where $\oint$ denotes the contour integration around the origin on the complexified $x$ plane and
        \eq{
            T&:=\sum\limits_{n} L_n x^{-n-2}-\sum\limits_{n} (n+1)yM_{n-1}x^{-n-2},	\label{T_modes}\\
            M&:=\sum\limits_{n} M_n x^{-n-2}.	\label{M_modes}
        }
        From the algebra \eqref{eq: BMSalg} we have the following OPEs between the currents,
        \eq{
            \begin{aligned}
                T(x',y') T(x,y) & \sim \frac{c_L}{2(x'-x)^4}+ \frac{2T(x,y)}{(x'-x)^2}+\frac{\partial_x T(x,y)}{x'-x}  \\
                &\qquad-\frac{2c_M(y'-y)}{(x'-x)^5}- \frac{4(y'-y) M(x,y)}{(x'-x)^3}- \frac{(y'-y)\partial_y T(x,y)}{(x'-x)^2}, \\[3pt]
                T(x',y') M(x,y) & \sim \frac{c_M}{2 (x'-x)^4}+\frac{2M(x,y)}{(x'-x)^2}+\frac{\partial_x M(x,y)}{x'-x}, \\[6pt]
                M(x',y') M(x,y) & \sim 0.
            \end{aligned}
            \label{BMSope}
        }
        Under the transformation \eqref{eq: BMSplane}, the transformation laws of the stress tensors are given by
        \eq{\label{eq: translaw}
            \widetilde{M}(x)&=f^{\prime 2} M\left(\widetilde{x} \right)+\frac{c_{M}}{12}\{f, x\},\\
            \widetilde{T}(x, y)&=f^{\prime 2} T\left(\widetilde{x}, \widetilde{y}\right)+2 f^{\prime } \big(g^{\prime}+y f''\big) M\left(\widetilde{x}\right)+\frac{c_{L}}{12}\{f, x\}+ \frac{c_{M}}{12} \Big(y\partial_x\{f,\,x\}+ f'^2 \frac{\partial^3 g}{\partial f^3}\Big),\nonumber
        }
        where $\{\,,\,\}$ denotes the usual Schwarzian derivative, and the last term is the so-called BMS Schwarzian derivative,
        \eq{
            \{f,\,x\}&=\frac{{f}^{\prime \prime \prime}}{{f}^{\prime}}-\frac{3}{2}\left(\frac{{f}^{\prime \prime}}{{f}^{\prime}}\right)^{2},\\
            f'^2 \frac{\partial^3 g}{\partial f^3} &=f^{\prime-1}\left(g^{\prime\prime\prime}-g^{\prime} \frac{{f}^{\prime \prime \prime}}{{f}^{\prime}}-3 f^{\prime \prime}\left(\frac{g^{\prime}}{f^{\prime}}\right)^{\prime}\right).
        }
        For later convenience,  we can define
        \begin{equation}
            T_0(x)=T(x,0)=T(x,y)-y\partial_x M(x),
        \end{equation}
        which leads to the conservation law of the stress tensor
        \begin{equation}
            \partial_y T_0=\partial_yT-\partial_x M=0,\quad \partial_y M=0.
        \end{equation}
        The charges $L_n$ can related to $T_0$ directly,
        \eq{\label{modes2}
            L_n=\frac{1}{2\pi i}\oint  x^{n+1} T_0dx,\qquad T_0:=\sum\limits_{n} L_n x^{-n-2}.
        }
        The transformation law of the newly defined operator $T_0$ reads,
        \begin{equation}\label{eq:transt0}
    \widetilde{T}_0(x)=f^{\prime 2} T_0\left(\widetilde{x}\right)+2 f^{\prime } g^{\prime} M\left(\widetilde{x}\right)+f^{\prime 2} g \partial_{\widetilde{x}}M\left(\widetilde{x}\right)+\frac{c_{L}}{12}\{f, x\}+ \frac{c_{M}}{12} f'^2 \frac{\partial^3 g}{\partial f^3}.
        \end{equation}

        The operators in the highest weight representation can be organized into  primaries and their descendants.
        The primary operators at the origin are labeled by the eigenvalues $(\Delta, \xi)$ of $(L_0, M_0)$
        \begin{equation}
            [L_0,O]=\Delta O,\qquad [M_0,O]=\xi O,
        \end{equation}
        where $\Delta$ and $\xi$ are referred to as the conformal weight and the boost charge  respectively.
        They obey the highest weight conditions
        \begin{equation}
            [L_n,O]=0,\qquad [M_n,O]=0, \qquad n>0.
        \end{equation}
        Acting $L_{-n},M_{-n}$ with $n>0$ successively on the primaries, we  get their descendants.
        This definition can be generalized to the multiplet.
        A highest-weight primary multiplet $\bO$ with rank $r$ is defined by
        \eq{
            [L_0,\,O_{a}]&=\Delta O_{a},\qquad [M_0,O_{a}]=({{\bxi}} O)_a,\quad a=0,\cdots r-1\nonumber\\
            [L_n,\,O_{a}]&=0,\quad  [M_n,O_{a}]=0,  \quad n>0
            ,\label{multiplet}
        }
        where $O_{a}$ denotes the $a$-th component of the multiplet $\bO$, and ${{\bxi}}$ is a Jordan cell with rank $r$ and diagonal component $\xi$,
        \begin{equation}\label{jordan}
            \bxi=
            \begin{pmatrix}
                \xi& & & \\
                1& \xi& & \\
                & \ddots&\ddots &\\
                & & 1& \xi\\
            \end{pmatrix}_{r\times r}.
        \end{equation}
        In particular, the finite transformation law for the multiplet can be obtained as
        \begin{equation}\label{mftran}
            \tilde{O}_{a}(\tilde{x}, \tilde{y})=\sum_{k=0}^{a}\frac{1}{k!}|f'|^{-\Delta}\,\partial_{\xi}^{k}e^{-\xi\frac{g'+yf''}{f'}}\,O_{a-k}(x,y).
        \end{equation}
        For singlet, this reduces to
        \begin{align}
            \tilde{O}(\tilde{x}, \tilde{y})=f'^\Delta e^{-\xi\frac{g'+yf''}{f'}}\,O(x,y). \label{eq: Otrans}
        \end{align}

        The vacuum is invariant under the global symmetry algebra spanned by $\{L_{0,\pm1},M_{0,\pm1}\}$, leading to the correlation functions for singlet operator in the highest weight representation
        \begin{align}
            G_2=\Big\langle \prod_{i=1}^2O_i(x_i,y_i)\Big\rangle &= \frac{d \,\delta_{\Delta_1,\Delta_2}\delta_{\xi_1,\xi_2}}{|x_{12}|^{2\Delta_1}}e^{-2\xi_1\frac{ y_{12}}{x_{12}}}, \label{eq: 2pt} \\
            G_3=\Big\langle \prod_{i=1}^3O_i(x_i,y_i)\Big\rangle  &= \frac{c_{123}}{|x_{12}|^{-\Delta_{123}}|x_{23}|^{\Delta_{231}}|x_{31}|^{\Delta_{312}}}e^{\xi_{123}\frac{y_{12}}{x_{12}}}e^{-\xi_{312}\frac{y_{31}}{x_{31}}}e^{-\xi_{231}\frac{y_{23}}{x_{23}}}, \label{eq: 3pt}
        \end{align}
        where $d$ is the normalization factor of the 2-point function, $c_{123}$ is the coefficient of 3-point function which encodes dynamical information, and
        \begin{equation}
            x_{ij}\equiv x_i-x_j,\ \ y_{ij}\equiv y_i-y_j,\ \ \Delta_{ijk}\equiv\Delta_i+\Delta_j-\Delta_k,\ \ \xi_{ijk}\equiv\xi_i+\xi_j-\xi_k.
        \end{equation}
        The 4-point functions of singlet quasi-primary operators can be determined up to an arbitrary function of the cross ratios,
        \begin{equation}
            G_4=\Big\langle \prod_{i=1}^4O_i(x_i,y_i)\Big\rangle =\prod_{i,j}|x_{ij}|^{\sum_{k=1}^4 -\Delta_{ijk}/3}e^{-\frac{y_{ij}}{x_{ij}}\sum_{k=1}^{4}\xi_{ijk}/3}\mathcal{G}(x,y),
        \end{equation}
        where the indices $i=1,2,3,4$ label the external operators $O_i$, $\mathcal{G}(x,y)$ is called the stripped 4-point function and $x$ and $y$ are the cross ratios,
        \begin{equation}
            \label{eq: cross ratio}x\equiv\frac{x_{12}x_{34}}{x_{13}x_{24}},\qquad \frac{y}{x}\equiv\frac{y_{12}}{x_{12}}+\frac{y_{34}}{x_{34}}-\frac{y_{13}}{x_{13}}-\frac{y_{24}}{x_{24}}.
        \end{equation}
        For the stripped 4-point function of identical external operators, its global block expansion is
        \begin{equation} \label{4pointexpansion}
            \mathcal{G}(x,y)=\sum_{O_r,\xi_r\neq0}\frac{1}{d_r}g|_{\xi\neq0}+\sum_{O_r,\xi_r=0}\frac{1}{d_r}g|_{\xi=0},
        \end{equation}
        where $g$ is the global block.
        We first consider the expansion in the s-channel.
        The propagating operators with rank-$r$ and non-zero boost charge $\xi_r\neq0$ has the contribution
        \begin{equation} \label{4pointexpansion_nonzero}
            g|_{\xi\neq0}=\sum_{p=0}^{r-1}\frac{1}{p!}\sum_{a,b|a+b+p+1=r}c_ac_b\partial_{\xi_r}^pg^{(0)}_{\Delta_r,\xi_r},
        \end{equation}
        where $g^{(0)}_{\Delta_r,\xi_r}$ is the global C/G conformal block for a propagating singlet with $\xi\neq0$,
        \begin{equation}
            g^{(0)}_{\Delta_r,\xi_r}=2^{2\Delta_r-2}x^{\Delta_r}(1+\sqrt{1-x})^{2-2\Delta_r}e^{\frac{\xi_r y}{x\sqrt{1-x}}}(1-x)^{-1/2}.  \label{singletblock}
        \end{equation}
        The contribution of propagating operators with zero boost charge $\xi_r=0$ can be expressed as
        \begin{equation} \label{xi=0block1}
            g|_{\xi=0}\ =\sum_{i,j|i+j+p=r-1}c_ic_jg_{ij}=\sum_{p=0}^{r-1}\sum_{i=0}^{r-1-p}c_ic_jg_{ij}|_{j=r-1-p-i},
        \end{equation}
        where
        \begin{equation} \label{xi=0block2}
            g_{ij}=\frac{1}{p!}x^\Delta q^p \sum_{a=0}^{\min(j,r-1-j)}x^af_a\ _2F_1(\Delta+a,\Delta+a+p,2\Delta+2a,x).
        \end{equation}
        The coefficients $f_a$ can be written into a double summation,
        \begin{equation}
            f_{a}=\frac{\sin (\pi  \Delta ) \Gamma
               (a+2 \Delta -1) \sin (\pi
               (\Delta +p)) \Gamma (a+p+\Delta
               )}{\pi ^{3/2} \Gamma (a+1)
               \Gamma \left(a+\Delta
               -\frac{1}{2}\right)}\sum_{b,c=0}^a
            f_{bc}^{(1)}f_{bc}^{(2)},
        \end{equation}
        where
        \begin{align}
            f_{bc}^{(1)} & :=\frac{(-1)^{-a+b+c} \Gamma
               (-c-\Delta +1) 2^{-2 a-b-c-2
               \Delta +2} \Gamma (b+c+2 \Delta
               -1) \Gamma (-b-p-\Delta
               +1)}{\Gamma (b+c+1)}, \nonumber \\
            f_{bc}^{(2)} & :=
            \, _4\widetilde{F}_3(1,-a,-b-c,a+2
               \Delta -1;1-b,1-c,2 \Delta -1;1).
        \end{align}
        The first few order terms are listed explicitly as
        \begin{equation}
            f_0=1,\ \ f_1=-2\Delta-p,\ \ f_2=\frac{(3+8\Delta)(3+2p(2+p)+11\Delta+9p\Delta+8\Delta^2)}{16+32\Delta}.
        \end{equation}
        The following facts will be useful in subsequent discussions.
        Especially, we consider the operators with weight $\Delta$ and charge $\xi$ located as follows, $\langle O(\infty,0)O(1,0)O(x,y)O(0,0)\rangle$.
        The contribution of the $\xi_r\neq0$ singlet is
        \begin{equation}
            g(x,y)=2^{2\Delta_r-2}x^{\Delta_r-2\Delta}(1+\sqrt{1-x})^{2-2\Delta_r}e^{-2\xi\frac{y}{x}+\frac{\xi_r y}{x\sqrt{1-x}}}(1-x)^{-1/2},
        \end{equation}
        and the contribution of the $\xi_r=0$ singlet is
        \begin{equation}
            g(x,y)=x^{\Delta_r-2\Delta}e^{-2\xi\frac{y}{x}}\ _2F_1(\Delta,\Delta,2\Delta,x).
        \end{equation}
        These global blocks are normalized as $x^{-2\Delta+\Delta_r}e^{(-2\xi+\xi_r)\frac{y}{x}}$ in the $x\rightarrow0$ limit.
        The contribution from the vacuum is the kinematic factor $x^{-2\Delta}e^{-2\xi\frac{y}{x}}$.
        The $t$-channel expansion can be obtained by the replacement,
        \begin{equation}
            x\rightarrow 1-x,\ \ y\rightarrow -y.
        \end{equation}

    \section{Conformal blocks in 2d C/G CFTs} \label{sec: bms block}
        In this section, we deal with the C/G conformal blocks organized by the entire C/G algebra instead of the global C/G algebra, in the large central charge limit.
        This paper focuses on the scenario where the propagating operator is a singlet, serving as the starting point for block calculations.

        One can define the projector $|\mathbf{O}|$ onto the family of $\mathbf{O}$ containing all the descendant operators related as
        \begin{equation}|\mathbf{O}|:=\sum_{\alpha,\beta}(N^{-1})_{\alpha\beta} |\alpha\rangle\langle\beta|,
        \end{equation}
        where the summation of $\alpha$ and $\beta$ is over all the descendent operators of $\mathbf{O}$, and
        $N_{\alpha\beta}=\langle\alpha|\beta\rangle$ is the inner-product matrix.
        We should also drop all the null states in this projector.
        The identity operator can be decomposed into the summation over all the projectors
        \begin{equation}
            1=\sum_{\mathbf{O}}|\mathbf{O}|.
        \end{equation}
        Here comes the following decomposition of a 4-point function
        \begin{equation}
            \langle O(\infty,0)O(1,0)O(x,y)O(0,0)\rangle=\sum_{\{\mathbf{O}\}}g_\mathbf{O}(x,y),
        \end{equation}
        where $g_\mathbf{O}(x)$ is the contribution from each operator family
        \begin{equation}
            g_\mathbf{O}(x,y)=\langle O(\infty,0)O(1,0)|\mathbf{O}|O(x,y)O(0,0)\rangle=
            \begin{tikzpicture}[baseline={([yshift = -0.5ex]current bounding box.center)}]
                \draw[thick] (-1.5,1)--(-0.8,0);
                \draw[thick] (-1.5,-1)--(-0.8,0);
                \draw[] (-0.8,0)--(0.8,0);
                \draw[thick] (1.5,1)--(0.8,0);
                \draw[thick] (1.5,-1)--(0.8,0);
                \node[above] at (0,0) {$\mathbf{O}$};
                \node[above] at (-1.5,1) {$O(\infty,0)$};
                \node[below] at (-1.5,-1) {$O(1,0)$};
                \node[below] at (1.5,-1) {$O(x,y)$};
                \node[above] at (1.5,1) {$O(0,0)$};
            \end{tikzpicture}
             \label{blockform}.
        \end{equation}
        Note that it contains the C/G conformal block together with the coefficient related to the 3-point coefficients.
        In the following, we consider the C/G blocks in the large central charge limit.
        The theory involves two central charges $c_L$ and $c_M$, so we must carefully examine which of these serves as the expansion parameter.

        \subsection{Light type}
            We first examine the case where all external operators are light, meaning their charges $\xi$ and weights $\Delta$ are $O(1)$ compared to the suitably chosen central charge.
            The conformal blocks and correlation functions involving only light operators are called the \textit{light type}.

        \subsubsection*{Large $c_M$ limit}
            First, we consider the four-point function in the following limit
            \begin{tcb}[Large $c_M$ limit]
                \begin{equation}\label{largecm}
                    c_M\rightarrow\infty,\ \ \text{with fiexed }\Delta,\xi,\Delta_r,\xi_r,\ \ \text{but }\frac{c_L}{c_M^{2}}\rightarrow0.
                \end{equation}
            \end{tcb}
            This condition encompasses several cases including $c_L\sim O(1)$, $c_L\sim O(c_M)$, and so forth.
            Under this limit, the C/G conformal block involving four identical external light operators reduces to the global C/G block, as we can see as follows.

            We denote the projector for a specific family at level $ n $ as
            \begin{equation}
                |O|_{n}:=\sum_{n=n_1m_1+\cdots+n_2m_2+\cdots} L_{-m_1}^{n_1}\cdots M_{-m_2}^{n_2}\cdots|\Delta_r,\xi_r\rangle\langle\Delta_r,\xi_r|\cdots M_{m_2}^{n_2}\cdots L_{m_1}^{n_1}(N^{-1})_{m_i,n_i},
            \end{equation}
            where $N^{-1}$ is the inverse matrix of the inner product at level $n$.
            Due to the fact that the descendant states at different levels are orthogonal, the projector can be decomposed into
            \begin{equation}
                |\mathbf{O}|=\sum_{n}|O|_n.
            \end{equation}
            We observe that the inner product matrix at each level $ n $ exhibits an approximate block-diagonal structure.
            The basis elements can be systematically labeled based on the total count of $ L_{-n} $ and $ M_{-n} $ operators with $ n \geq 2 $.
            When these basis elements are arranged in ascending order according to this count, the corresponding inner product matrix demonstrates a behavior characterized by its block-diagonal approximation
            \begin{align}
                \begin{pmatrix}
                    O(1) & O(1)  & \cdots & O(1)  \\
                    O(1)  & O(c_M) & \cdots &O(c_M) \\
                    \vdots & \vdots & \ddots & \vdots \\
                    O(1)  &O(c_M)& \cdots &O(c^n_M )
                \end{pmatrix}.
            \end{align}
            Then, the non-diagonal blocks of the inverse matrix exhibit a magnitude of at least $O(1/c_M)$.
            At leading order approximation, the projector can be decomposed into
            \begin{equation}
                |O|\approx\sum_{\{k\}}|O|_{k_i},
            \end{equation}
            where
            \begin{equation}
                |O|_{k_i}:= \sum_{k_i=m_i+n_i}L_{-1}^{n_1}\cdots M_{-1}^{m_1}\cdots|\Delta_r,\xi_r\rangle\langle\Delta_r,\xi_r|\cdots M_{1}^{m_1}\cdots L_{1}^{n_1}(N^{-1})_{m_i,n_i}.
            \end{equation}
            The vector $ k $ serves to label the number of generators present at each respective level within the system.
            To illustrate this concept, we provide examples of contributions at the initial orders below,
            \begin{itemize}
                \item $k=(1,0,0,\cdots)$, the basis are $L_{-1}|O\rangle, M_{-1}|O\rangle$.
                \item $k=(2,0,0,\cdots)$, the basis are $L_{-1}L_{-1}|O\rangle, L_{-1}M_{-1}|O\rangle,M_{-1}M_{-1}|O\rangle$.
                \item $k=(0,1,0,\cdots)$, the basis are $L_{-2}|O\rangle, M_{-2}|O\rangle$.
            \end{itemize}
            The summation encompasses all possible values of the vector $ k $.
            When considering the large $ c_M $ expansion, the inner product matrix $N_k$ can be further represented as a tensor product,
            \begin{equation}
                N_{k_1+k_2}=N_{k_1}\otimes N_{k_2},\ \ \text{if $k_1$,$k_2$ are orthogonal.}
            \end{equation}
            Sequently, the analysis of the general case $ N_k $ simplifies to the analysis of $ N_{(0,0,\cdots,a,0,0,\cdots)} $.
            This specific configuration encompasses $ a + 1 $ distinct states, and the inner products between these states can be computed through straightforward procedures, leading to
                \begin{equation}
                    {N_{(0,0,\cdots,a,0,0,\cdots)}}_{ij}=\left\{
                        \begin{array}{lll}
                            & c_L^{a-i-j+2}c_M^{i+j-2} & (i+j\leq a+2), \\[6pt]
                            &  0\ \ \ \  \ \ & (\text{otherwise}).
                        \end{array}
                    \right.
            \end{equation}
            The indices $i, j$ denote the basis elements based on the count of $M_n$ generators.
            The inverse of the above inner product matrix has the following form
            \begin{align}
                {N_{(0,0,\cdots,a,0,0,\cdots)}}^{-1}=   \begin{pmatrix}
                    0 & 0 & \cdots & \frac{1}{c_M^a} \\
                    0 & 0 & \frac{1}{c_M^a} &-\frac{c_L}{c_M^{a+1}} \\
                    \vdots & \vdots & \vdots & \vdots \\
                    \frac{1}{c_M^a} &-\frac{c_L}{c_M^{a+1}}& \cdots &0
                \end{pmatrix}.
            \end{align}
            With the inverse matrix provided, it can be observed that the corresponding contributions are suppressed by factors of $\frac{1}{c_M^a}$ and $\frac{c_L}{c_M^{a+1}}$.
            These suppression factors, in turn, establish the final condition outlined in eq.\,\eqref{largecm}.

            We also examine the contributions arising from three-point functions.
            These contributions are proportional to the parameters $\Delta, \xi, \Delta_r, \xi_r$.
            A detailed derivation of their explicit expressions will be provided in the subsequent subsection.
            In the context of the large $c_M$ limit, as defined by eq.\,\eqref{largecm}, all contributions stemming from higher descendant states generated by the operators $L_{-n}, M_{-n} $where $n \geq 2$ are suppressed by a factor of at least $1/c_M$.
            Consequently, the projector simplifies to

            \begin{equation}
                |\mathbf{O}|\approx\sum_{m,n}L_{-1}^mM_{-1}^n|O\rangle\langle O|M_1^nL_1^m(N^{-1})_{mn}+O\Big(\frac{1}{c_M}\Big).
            \end{equation}
            The leading term gives the global C/G blocks for the propagating operator $O$, when inserted into the four point functions.
        \subsubsection*{Large $c_L$ limit}
            We next proceed to analyze the limit $c_L\to\infty$.
            It is important to note that the pattern $c_L \sim c_M\to\infty$ has already been addressed in the preceding discussion.
            We will now focus on the following case.
            \begin{tcb}[Large $c_L$ limit]
                \begin{equation}\label{largecl}
                    c_L\rightarrow\infty,\ \ \text{with fiexed }\Delta,\xi,\Delta_r,\xi_r, c_M.
                \end{equation}
            \end{tcb}
            The contributions arising from three-point functions remain same with those in the previous case, so our primary focus shifts to the contributions derived from the inner product.
            Specifically, we examine any state that belongs to the basis
            \begin{equation}
                L_{-1}^{0}L_{-2}^{n_2}\cdots|O\rangle
            \end{equation}
            Note that these states are orthogonal to all other states to the leading order in the inverse central charge expansion, specifically in terms of $1/c_L$.
            Their norms exhibit a proportionality to $c_L^a$, where $a:=\sum_{i=2}^\infty n_i$.
            This scaling behavior results in their contributions being significantly suppressed, so we may now focus on the remaining states, which form the basis under consideration,
            \begin{equation}
                L_{-1}^{q}M_{-1}^{m_1}M_{-2}^{m_2}\cdots|O\rangle.
            \end{equation}
            We first consider the states of the form $ M_{-1}^{m_1}M_{-2}^{m_2}\cdots|O\rangle $.
            These states exhibit a non-vanishing inner product exclusively with the single state $L_{-1}^{q}|O\rangle$, for which there exist $n-1$ null states at level $n$.
            Upon subtracting these null states from the total set of states at level $ n $, the remaining states are
            \begin{equation}
                L_{-1}^{q}M_{-1}^{p}|O\rangle.
            \end{equation}
            This process subsequently yields the global C/G blocks.
            Consequently, the projector in the context of the large central charge limit \eqref{largecl} takes the following form,
            \begin{equation}
                |\mathbf{O}|\approx \sum_{m,n}L_{-1}^mM_{-1}^n|O\rangle\langle O|M_1^nL_1^m(N^{-1})_{mn}+O(\frac{1}{c_L}).
            \end{equation}
            To summarize, as long as when the large central charge limits \eqref{largecm} and \eqref{largecl} are considered, the C/G conformal blocks of light type become the global blocks.

        \subsection{Heavy-light type}
            In this subsection, we examine another case where two external operators are heavy, in the sense that their weights $\Delta$ and charges $\xi$ being linear order in the large central charge, while the remaining two operators are light.
            This configuration gives rise to the \textit{heavy-light type} of the C/G conformal block, so-named because of its external operator contents.
            In the subsequent discussion, we focus on the large $c_M$ limit and do not elaborate on $c_L$ limit since it is hard to control.
            Note that this choice is not only for mere convenience but physically plausible, given the holographically dual picture of 3d Einstein gravity in semiclassical regime $G\to 0$ where the central charges of asymptotic symmetry algebra are given by $c_L=0$, $c_M = \frac{3}{G}$.

            We define our notation by examining the following four-point function,
            \begin{equation}
                \langle O_H(\infty,0)O_H(1,0)O_L(x,y)O_L(0,0)\rangle,
            \end{equation}
            with the weights and charges of the operators denoted as
            \begin{equation}
                \Delta_H=H,\ \ \xi_H=\Xi,\ \ \Delta_L=\Delta,\ \ \xi_L=\xi,
            \end{equation}
            and $\Delta_r,\xi_r$ are the weight and charge of the propagating operator considered, which are fixed under the large central charge limit.

            Since the weights and charges of the heavy external operators are not fixed in the large central charge limit, contributions from three-point functions involving higher descendant states become significant.
            These contributions can compare to the contribution from the large central charges in the inverse inner product matrix.
            To properly account for this, we will examine two pairs of three-point functions that emerge after inserting the projector.
            The one where the heavy operator is involved is
            \begin{equation}\label{3ptmodes}
                \langle O_H(\infty,0)O_H(1,0)L_{-1}^{n_1}\cdots M_{-1}^{m_1}\cdots|O\rangle.
            \end{equation}
            The other one involves light operators without dependence on the heavy weights and/or charges.
            So we aim to investigate the dependence on the weights, charges, and central charges in eq.\,\eqref{3ptmodes}.
            Equivalently, utilizing equations \eqref{modes1} and \eqref{modes2}, we may consider
            \begin{equation}
                \langle O_H(\infty,0)O_H(1,0)T_0(x_1)\cdots M(x_2) \cdots O(0,0)\rangle,
            \end{equation}
            which can be further structured from the stress tensor OPE and resulting Wald identity,
            \eq{&\langle O_H(\infty,0)O_H(1,0)T_0(x_1)\cdots M(x_2) \cdots O(0,0)\rangle \\[4pt]
                &\qquad\sim\,\sum_i\langle O_H(\infty,0)O_H(1,0)T_0(x_i) O(0,0)\rangle+\sum_i\langle O_H(\infty,0)O_H(1,0) M(x_i) O(0,0)\rangle+O(c_L,c_M), \nonumber
            }
            where $O(c_L,c_M)$ is the central charge part, which will not contribute to eq.\,\eqref{3ptmodes} due to the vanishing residue.
            So the fundamental parts are
            \begin{equation}\label{eq: HHTO}
                \langle O_H(\infty,0)O_H(1,0)T_0(x) O(0,0)\rangle=c_{AAO}\Big(\frac{\Delta_r}{x^2(1-x)}+\frac{H}{(x-1)^2}\Big),
            \end{equation}
            \begin{equation}\label{eq: HHMO}
                \langle O_H(\infty,0)O_H(1,0)M(x) O(0,0)\rangle=c_{AAO}\Big(\frac{(2x-1)\xi_r}{x^2(1-x)}+\frac{\Xi}{(x-1)^2}\Big).
            \end{equation}
            In the following discussion, we first consider different cases where some weights and/or charges are not fixed under the large $c_M$ limit, then find the C/G conformal transformation to cancel these dependence on the large weights and/or charges by the anomalous terms in the stress tensors.
        \subsubsection*{Large $\Xi$}
            We start with the simplest case with large $\Xi$ with the following limit
            \begin{tcb}[Large $\Xi$]
                \begin{equation}\label{eq: large xi}
                    c_M\rightarrow\infty,\ \ c_L,H,\Delta,\xi,\Delta_r,\xi_r\sim O(1),\ \ \frac{\Xi}{c_M}\sim O(1).
                \end{equation}
            \end{tcb}
            We want to find a C/G conformal transformation
            \begin{equation}
                x=f(w),\ \ y=f'(w)z+g(w),
            \end{equation}
            to cancel the $\Xi$ dependence in eq.\,\eqref{eq: HHMO}, so that
            \begin{equation}
                \lim_{c_M\to\infty} \partial_\Xi \langle O_HO_HM(w)O\rangle=0.
            \end{equation}
            This condition leads to the differential equation on the C/G conformal transformation $f(w)$, by considering the transformation law \eqref{eq: Otrans}, \eqref{eq: translaw}
            \begin{equation}
                f'(w)^2\frac{\Xi}{(f(w)-1)^2}+\frac{c_M}{12}\{f(w),w\}=0.
            \end{equation}
            The general solution contains three integration constants, which correspond to global transformations that prevent anomalies in the stress tensor $M$.
            These constants can be employed to determine the positions of the $O_H$ and $O$ operators.
            For simplicity, we select $O(0, 0)$, $O_H(1, 0)$, and $O_H(\infty, 0)$.
            The resulting solution is as follows,
            \begin{equation}
                \label{eq: f solution}    x=1-(1-w)^{\frac{1}{\alpha}},
            \end{equation}
            with $\alpha=\pm\sqrt{1-\frac{24\Xi}{c_M}}$.
            We will choose the positive one without loss of generality.
            This transformation will give additional terms in eq.\,\eqref{eq: HHTO}
            \begin{equation}\label{eq: ooto term}
                \langle O_HO_HT_0(w)O\rangle= \langle O_HO_H\big(f'^2T_0(x)+ 2 f^{\prime } g^{\prime} M\left(x\right)+f^{\prime 2} g \partial_{x}M\left(x\right)\big)O\rangle+\frac{c_{L}}{12}\{f, w\}+ \frac{c_{M}}{12} f'^2 \frac{\partial^3 g}{\partial f^3}
            \end{equation}
            Since $\langle O_HO_HM(x)O\rangle$ has term linear in $\Xi$ and $\{f,w\}$ is proportional to $\frac{\Xi}{c_M}$ which is $O(1)$, we can simply choose $g(w)=0$ so that
            \begin{equation}
                \lim_{c_M\rightarrow \infty}\partial_\Xi\langle O_HO_HT_0(w)O\rangle=0.
            \end{equation}
            So we conclude that in this case, the C/G conformal transformation
            \begin{equation}\label{trans1}
                1-w=(1-x)^\alpha,\ \ z=\alpha(1-x)^{\alpha-1}y,\ \ \alpha=\sqrt{1-\frac{24\Xi}{c_M}},
            \end{equation}
            can cancel the large $\Xi$ dependence in eq.\,\eqref{3ptmodes}.
            It possesses fixed points at $(0,0)$, $(1,0)$, and $(\infty,0)$.
            The C/G conformal algebra and highest weight states remain preserved.
            A branch cut is introduced on the w-plane extending from 1 to $\infty$, rendering the Laurent expansion at infinity invalid and causing the adjoint to be ill-defined.
            Despite this, out-states and in-states in the $(w,z)$ coordinates can be defined conventionally, though they are not adjoint to one another.

            Since the highest weight family is mapped to the family marked with the same primary operator under the C/G conformal transformation, the projector in different coordinates collect same contributions, while the four point functions are up to Jacobian due to the transformation law \eqref{eq: Otrans},
            \begin{equation}
                g_{\mathbf{O}}(x,y)=\langle O_H(\infty,0)O_H(1,0)|\mathbf{O}|O_L(x,y)O_L(0,0)\rangle=J_{O_L(x,y)}J_{O_L(0,0)}\widetilde{g}_{\mathbf{O}}(w,z),
            \end{equation}
            where $J$ is the Jacobian factor
            \begin{equation}\label{eq: jacobian}
                J_O=f'^2e^{\xi\frac{g'+yf''}{f'}}.
            \end{equation}
            Note that $(x,y)$ is the flat plane coordinate, while $(w,z)$ is not.
            Now the calculation of the full blocks in $(x,y)$ coordinate reduces to the calculation of the global blocks in $(w,z)$ coordinate, since the inner product in $(w,z)$ coordinate has the same form when considered in flat coordinate and the three point functions has no dependence on the heavy charge $\Xi$ as discussed above.
            Besides, when we consider the global blocks $g_{\text{global}}$ in $(w,z)$ coordinate, the actions of $L_{-1},M_{-1}$ has the same form as the flat one up to a Jacobian factor of the propagating operator
            \begin{equation}
                \widetilde{g}_\mathbf{O}(w,z)=J_r^{-1}g_{\text{global}}(w,z)+O\Big(\frac{1}{c_M}\Big).
            \end{equation}
            So in the setup \eqref{eq: large xi},
            \begin{equation}
                \lim_{c_M\rightarrow\infty}g(x,y)=(1-w)^{\Delta(1-\frac{1}{\alpha})}\alpha^{2\Delta-\Delta_r}e^{\xi\frac{z(\alpha-1)}{(w-1)\alpha}}g_{\text{global}}(w,z). \label{eq: general block to global block 1}
            \end{equation}
            Compared the CFT results in Ref.\,\cite{Fitzpatrick:2015zha}, it shares a similar form but incorporates three distinct modifications.
            The parameter $\alpha$ differs through the substitutions $H\to\Xi, c\to c_M$.
            Also, an additional exponential factor $e^{\xi\frac{z(\alpha-1)}{(w-1)\alpha}}$ arises from the Jacobian \eqref{eq: jacobian}.
            The global block also varies since there exist two types of global C/G blocks for singlets with $\xi_r=0$ and $\xi_r\neq0$.
            In the $\xi_r=0$ case, the global C/G block matches the global $SL(2,R)$ block up to a normalization factor, which stems from the kinematic term $e^{-2\xi\frac{z}{w}}$.

            In particular, the vacuum block which we will use later reads,
            \eq{\lim_{c_M\rightarrow\infty}g_{vac}(x,y)&=(1-w)^{\Delta(1-\frac{1}{\alpha})}(\frac{w}{\alpha})^{-2\Delta}e^{\xi\frac{z(\alpha-1)}{(w-1)\alpha}}e^{-2\xi\frac{z}{w}} \\
                &=(1-x)^{(\alpha -1) \Delta }
                \left(\frac{1-(1-x)^{\alpha
                }}{\alpha }\right)^{-2
                \Delta } \exp
                \left(-\frac{\xi  y
                \left(\alpha +\alpha
                (1-x)^{\alpha
                }+(1-x)^{\alpha
                }-1\right)}{(x-1)
                \left((1-x)^{\alpha
                }-1\right)}\right). \nonumber}
        \subsubsection*{Large $H$}
            We then consider an opposite limit with large $H$,
            \begin{tcb}[Large $H$]
                \begin{equation}\label{eq: large H}
                    c_M\rightarrow\infty,\ \ c_L,\Xi,\Delta,\xi,\Delta_r,\xi_r\sim O(1),\ \ \frac{H}{c_M}\sim O(1).
                \end{equation}
            \end{tcb}
            We want to find a C/G conformal transformation
            \begin{equation}
                x=f(w),\ \ y=f'(w)z+g(w),
            \end{equation}
            to cancel the $H$ dependence in the $\langle O_HO_HT_0O\rangle$ correlator \eqref{eq: HHTO}, so that
            \begin{equation}
                \label{eq: hhto cm}\lim_{c_M\to\infty} \partial_H \langle O_HO_HT_0(w)O\rangle=0.
            \end{equation}
            At the same time, this transformation may introduce $c_M$ dependence in the $\langle O_HO_HM(w)O\rangle$ correlator, by the transformation law \eqref{eq: translaw}.
            So we require
            \begin{equation}
                \{f,w\}=0,
            \end{equation}
            to avoid the linear $c_M$ term.
            We can further fix the point $(0,1,\infty)$ on the $x$ plane, leading to
            \begin{equation}
                x=f(w)=w.
            \end{equation}
            Now in the large $c_M$ limit, eq.\,\eqref{eq: hhto cm} reduces to
            \begin{equation}
                f'^2\frac{H}{(x-1)^2}+ \frac{c_{M}}{12} f'^2 \frac{\partial^3 g}{\partial f^3}=0.
            \end{equation}
            Similarly, the solution has three integration constants related to the global transformations.
            We choose
            \begin{equation}\label{trans2}
                z=y+\frac{12H}{c_M}(1-x)\log(1-x).
            \end{equation}
            Although $(0,0),(1,0)$ are the fixed point of this transformation, $(\infty,0)$ in the $(x,y)$ coordinate is mapped to $(\infty,\infty)$ in the $(w,z)$ coordinate.
            However, the cross ratios \eqref{eq: cross ratio} remain invariant.
            This transformation leads to the Jacobian
            \begin{equation}
                J=\exp\Big\{-\frac{12H}{c_M}\big(1+\log(1-x)\big)\xi\Big\}.
            \end{equation}
            Hence we have the full C/G blocks in eq.\,\eqref{eq: large H}
            \begin{equation}
                \lim_{c_M\rightarrow\infty}g(x,y)=(1-w)^{-\frac{12H \xi}{c_M}}e^{-\frac{12H}{c_M}(2\xi-\xi_r)}g_{\text{global}}(w,z), \label{eq: general block to global block 2}
            \end{equation}
            where the constant factor is important, which contributes to the normalization $x^{-2\Delta+\Delta_r}e^{(-2\xi+\xi_r)\frac{y}{x}}$ in the $(x,y)$ coordinate.

        \subsubsection*{Large $H,\Xi,c_L$}
            One more general case is the limit with large $H,\Xi,c_L$
            \begin{tcb}[Large {$H,\Xi,c_L$}]
                \begin{equation}\label{eq: large H Xi}
                    c_M\rightarrow\infty,\ \ \Delta,\xi,\Delta_r,\xi_r\sim O(1),\ \ \frac{H}{c_M},\frac{\Xi}{c_M},\frac{c_L}{c_M}\sim O(1).
                \end{equation}
            \end{tcb}
            We want to find a C/G conformal transformation
            \begin{equation}
                x=f(w),\ \ y=f'(w)z+g(w)
            \end{equation}
            to cancel all the $H,\Xi,c_L$ dependence in eqs.\,\eqref{eq: HHTO}, \eqref{eq: HHMO}, so that
            \begin{equation}
                \lim_{c_M\rightarrow\infty}     \partial_\Xi \langle O_HO_HM(w)O\rangle=0,\ \ \lim_{c_M\rightarrow\infty}\partial_{H,\Xi,c_L} \langle O_HO_HT_0(w)O\rangle=0.
            \end{equation}
            We first look into the $\langle O_HO_HM(w)O\rangle$ correlator, where the discussion is in analog to that in the large $\Xi$ case, leading to the same solution as \eqref{eq: f solution},
            \begin{equation}\label{eq: sol f}
                x=1-(1-w)^{\frac{1}{\alpha}}.
            \end{equation}
            Then we consider \eqref{eq: HHTO} with terms in eq.\,\eqref{eq: ooto term}, resulting the differential equation on the $g$-transformation,
            \begin{equation}
                \frac{H}{(x-1)^2}+(2g'(x)-\frac{c_L}{c_M})\frac{\Xi}{(x-1)^2}+g(x)\partial_x\frac{\Xi}{(x-1)^2}+\frac{c_M}{12}g'''(x)=0.
            \end{equation}
            The solution which keeps $(0,0)$ and $(1,0)$ invariant is
            \begin{equation}\label{eq: sol g}
                g(x)=\alpha Q(1-x)^\alpha\log(1-x),
            \end{equation}
            where $Q$ is the parameter depending to $c_L,c_M,H,\alpha(\Xi)$,
            \begin{equation}
                Q=\frac{(\alpha^2-1)c_L+24H}{2\alpha^2c_M}.
            \end{equation}
            This transformation can be seen as the composite transformation of \eqref{trans1}, \eqref{trans2}, with the additional contribution from $c_L$ which can be either small or large.
            In this case the Jacobian factor is
            \begin{equation}
                J=(1-w)^{\Delta(1-\frac{1}{\alpha})}\alpha^\Delta(1-w)^{-\xi\frac{Q}{\alpha}}\exp\Big\{\xi\big(- Q+\frac{z(\alpha-1)}{(w-1)\alpha}\big)\Big\},
            \end{equation}
            leading to the full C/G conformal blocks in the limit \eqref{eq: large H Xi},
            \begin{equation}
                \lim_{c_M\rightarrow\infty}g(x,y)=(1-w)^{\Delta(1-\frac{1}{\alpha})-\xi\frac{Q}{\alpha}}\alpha^{2\Delta-\Delta_r}e^{\xi\frac{z(\alpha-1)}{(w-1)\alpha}+Q(\xi_r-2\xi)}g_{\text{global}}(w,z). \label{eq: general block to global block 3}
            \end{equation}
            It is important to note that the discussion and results presented here remain valid even when some of the parameters $ c_L, H, \Xi $ are small.
            In such cases, the appropriate small parameter limits should be taken into account.
            Particularly, the vacuum block is expressed as follows,
                \begin{align}
                 \lim_{c_M\rightarrow\infty}   g_{vac}(x,y)&=(1-w)^{\Delta(1-\frac{1}{\alpha})-\xi\frac{Q}{\alpha}}(\frac{w}{\alpha})^{-2\Delta}e^{\xi\frac{z(\alpha-1)}{(w-1)\alpha}-2Q\xi}e^{-2\xi\frac{z}{w}}\nonumber\\
                    &=\left(\frac{1-(1-x)^{\alpha }}{\alpha
                       }\right)^{-2 \Delta }(1-x)^{(\alpha -1) \Delta +\alpha  \xi  Q
                       \left(\frac{2}{(1-x)^{\alpha }-1}+1\right)}\nonumber\\
                    &\qquad\qquad\qquad\times\exp \left(\frac{\xi  y \left(\alpha
                       \left(-\frac{2}{(1-x)^{\alpha
                       }-1}-1\right)-1\right)}{x-1}-2 \xi
                       Q\right).
                \end{align}
                The $t$-channel block has a simpler form which we will use later,
                \begin{equation}\label{tblock}
                   \lim_{c_M\rightarrow\infty} g_{vac}(1-x,-y)=\left(\frac{1-x^{\alpha }}{\alpha }\right)^{-2
                    \Delta } x^{(\alpha -1) \Delta
                    +\frac{\alpha  \xi  Q \left(x^{\alpha
                    }+1\right)}{x^{\alpha }-1}} \exp \left(-2
                    \xi  Q-\frac{\xi  y \left(\alpha +(\alpha
                    +1) x^{\alpha }-1\right)}{x
                    \left(1-x^{\alpha }\right)}\right).
                \end{equation}

        \subsection{Turn to the cylinder}
            In the preceding discussion, we focused on four-point functions on the $(x, y)$ plane.
            To clarify further, we transition to the cylinder described by $(u, \phi)$ coordinates.
            The transformation from the plane to the cylinder is
            \begin{equation}
                \phi=-i\log x,\ \ \ u=-i\frac{y}{x}, \label{eq: coordinate transf to cylinder with 2pi periodicity}
            \end{equation}
            with periodic identification $\phi\sim \phi+2\pi$.
            Then we can define another cylinder with $(u',\phi')$ coordinate related to the $(w,z)$ coordinate by the similar cylinder to plane map,
            \begin{equation}
                \phi'=-i\log w,\ \ \ u'=-i\frac{z}{w}.
            \end{equation}
            The key transformation to solve the C/G blocks in eqs.\,\eqref{eq: sol f}, \eqref{eq: sol g} introduce the branch cut runs from $1$ to infinity on the $w$ plane, with the deficit angle $2\pi(1-\alpha)$.

            Then we can consider the four-point function and the block expansion on the cylinder to the periodicity manifest.
            Now the key transformation \eqref{eq: sol f}, \eqref{eq: sol g} become
            \begin{equation}
                \phi'= \alpha\phi,\ \  u'=\alpha u+Q\phi,
            \end{equation}
            whose geometric interpretation is clear.
            It rescales and tildes the cylinder, making the new identification
            \begin{equation}
                (u',\phi')\sim(u'+2\pi\alpha Q,\phi'+2\pi\alpha).
            \end{equation}

    \section{Application: Excited state entanglement entropy in 2d C/G CFTs} \label{sec: bms entanglement entropy}
        In this section, we utilize the C/G conformal blocks derived in the previous section to compute the entanglement entropy of excited states in the context of large $c_M$, which corresponds holographically to 3D Einstein gravity in asymptotically flat spacetimes.
        We adhere to the same approach employed in the AdS$_3$/CFT$_2$ case, as outlined in Section \ref{sec: ee at large central charge in cft}.

        The entanglement entropy for an interval $A$ in 2d Carrollian/Galilean conformal field theories in the vacuum state can also be calculated by the twist operators \cite{Bagchi:2014iea}
        \begin{equation}
            S_A = -\lim_{n\to 1} \frac{1}{n-1} \log \Tr_A\rho_A^n,\quad \operatorname{Tr}\rho_A^n=\langle\mathcal{T}_n(x_a,y_a)\mathcal{T}_n(x_b,y_b)\rangle.
        \end{equation}
        The twist operators are inserted at the endpoints of the specified interval.
        Their weight and charge can be determined from the conformal transformation or the limiting procedure \cite{Bagchi:2014iea},
        \begin{equation}
            \Delta_n=\frac{c_L}{24}\Big(n-\frac{1}{n}\Big),\ \ \xi_n=\frac{c_M}{24}\Big(n-\frac{1}{n}\Big),
        \end{equation}
        depending on the central charges $c_L,c_M$.
        The entanglement entropy on the $(x,y)$ plane reads
        \begin{equation}
            S_A=\frac{c_L}{6}\log\frac{l_x}{\epsilon}+\frac{c_M l_y}{6l_x}, \label{eq: cylinder ee}
        \end{equation}
        where $l_x:=x_a-x_b,l_y:=y_a-y_b$.
        The cylinder result can be derived from the conformal transformation and/or direct Rindler method \cite{Jiang:2017ecm, Apolo:2020bld}
        \begin{equation}
            S_A=\frac{c_L}{6}\log(\frac{2}{\epsilon}\sin\frac{l_\phi}{2})+\frac{c_M}{12}l_u\cot\frac{l_\phi}{2},
        \end{equation}
        where $l_u,l_\phi$ are the length of the interval along the $u$ and $\phi$ directions on the cylinder with the identification $\phi\sim\phi+2\pi$.
        The result can be generalized to the  tilted cylinder with the identification $(u,\phi)\sim(u+2\pi\alpha Q,\phi+2\pi\alpha)$ by considering the transformation
        \begin{equation}
            u\to\alpha u+Q\phi,\ \ \phi\to \alpha\phi.
        \end{equation}
        In this case, the entanglement entropy reads,
        \begin{equation}\label{eq: tilt cyc ee}
            S_A=\frac{c_L}{6}\log\Big(\frac{2}{\alpha\epsilon}\sin\frac{ \alpha l_\phi }{2}\Big)+\frac{c_M}{6}\Big(Q+\frac{ \alpha (l_u-Ql_\phi)}{2}\cot\frac{\alpha l_\phi}{2}\Big).
        \end{equation}

        \subsection{Excited state entanglement entropy}
           We then consider the excited state in the highest weight representation
            \begin{equation}
                \label{eq: excited state}|\Delta_H,\xi_H\rangle:=O_H(0,0)|0\rangle,
            \end{equation}
            created by a heavy operator acting on the highest weight vacuum, with weight and charge $\Delta_H,\xi_H$.
            Then the trace of the $n$-th power of the reduced density matrix is
            \begin{equation}
                \operatorname{Tr}\rho_A^n=\langle O_H^n(0,0)\mathcal{T}_n(x,y)\mathcal{T}_n(1,0)O_H^n(\infty,0)\rangle,
            \end{equation}
             where we consider the interval $ A $ with endpoints at $ (1, 0) $ and $ (x, y) $ on the plane.
             This configuration corresponds to the heavy-light type four-point function in the $ n \rightarrow 1 $ limit, where the weights and charges are
            \begin{equation}
                H=n\Delta_H,\ \ \Xi=n\xi_H,\ \ \ \Delta=\Delta_n,\ \ \xi=\xi_n.
            \end{equation}
            Also we consider the theory with large central charge $c_M$,
            \begin{equation}
                c_M\gg 0,\ \ c_L\sim c_M\ \ (\text{or }c_L\sim O(1)).
            \end{equation}
            This aligns with our setup \eqref{eq: large H Xi}.
            Here, the aforementioned four-point function permits a $t$-channel block expansion, where the C/G blocks can be approximated by the heavy-light type block derived in the preceding section.
            Under the additional assumption of vacuum dominance, where the leading contribution in this block expansion stems from the vacuum block, we obtain the entanglement entropy of the excited state \eqref{eq: excited state} using the $t$-channel vacuum block \eqref{tblock},
            \eq{S_A&=-\log \frac{x}{\epsilon}\left(\frac{1}{12} (\alpha -1)
              c_L+\frac{\alpha  c_M Q
              \left(x^{\alpha }+1\right)}{12
              \left(x^{\alpha
              }-1\right)}\right)+\frac{1}{6} c_L
              \log \left(\frac{1-x^{\alpha }}{\alpha
              }\right)\nonumber\\
              &+\frac{1}{12}c_M \left(2
              Q+\frac{y \left(\alpha +(\alpha +1)
              x^{\alpha }-1\right)}{x \left(x^{\alpha
              }-1\right)}\right).
              \label{exee}
            }
            Note that the expansion is in $t$-channel, where the length of the interval is given by
            \begin{equation}
                l_x:=1-x,\ \ l_y:=y.
            \end{equation}
            Here, a few comments are in order.
            When $c_M = 0$, the result corresponds to that of chiral CFT$_2$, representing the leading-order term in the large central charge expansion.
            When $\alpha = 1$ and $Q = 0$, corresponding to a light background state, the expression reduces to the vacuum entanglement entropy, with corrections anticipated at the next order.

            Similarly, one can further generalize this discussion on the excited state entanglement entropy to the cylinder geometry with the identification $(u,\phi)\sim(u,\phi+2\pi)$, resulting
            \begin{equation}\label{eq: excited ee1}
                S_A=\frac{c_L}{6}\log\Big(\frac{2}{\alpha\epsilon}\sin\frac{ \alpha l_\phi }{2}\Big)+\frac{c_M}{6}\Big(Q+\frac{ \alpha (l_u-Ql_\phi)}{2}\cot\frac{\alpha l_\phi}{2}\Big).
            \end{equation}
            This expression is identical to  \eqref{eq: tilt cyc ee}, which represents the entanglement entropy of the vacuum state on a cylinder with the identification $(u,\phi)\sim(u+2\pi\alpha Q,\phi+2\pi\alpha)$.
            In other words, the entanglement entropy of an excited state on a cylinder with standard identification can be interpreted as the vacuum entanglement entropy on a cylinder with a non-trivial identification, determined by the charge and weight of the excited state.

            In the case where
            \begin{equation}
                \Xi>\frac{c_M}{24}
            \end{equation}
            the parameter $\alpha$ becomes purely imaginary, leading to the entanglement entropy as if in a thermal state
            \begin{equation}\label{eq: excited ee2}
                S_A=\frac{c_L}{6}\log\Big(\frac{2}{\alpha'\epsilon}\sinh\frac{ \alpha' l_\phi }{2}\Big)+\frac{c_M}{6}\Big(Q+\frac{ \alpha' (l_u-Ql_\phi)}{2}\coth\frac{\alpha' l_\phi}{2}\Big),
            \end{equation}
            where $\alpha'=i\alpha$.

        \subsection{Compare with the holographic entanglement entropy}
            This section compares the excited state entanglement entropy with the holographic entanglement entropy in constant mode solutions.
            On the bulk side, the dual theory corresponds to Einstein gravity in three-dimensional asymptotically flat spacetime.
            For flat holography, the holographic entanglement entropy is described by the swing surface proposal \cite{Jiang:2017ecm,Apolo:2020bld} (See also Appendix \ref{append: swing surface}).

            The central charge can be related to the gravitational constant   from the asymptotic symmetry analysis in Einstein gravity \cite{Barnich:2006av}
            \begin{equation}
                c_L=0,\ \ c_M=\frac{3}{G}.
            \end{equation}
            The solutions with the asymptotically flat boundary condition are organized in the Bondi gauge,
            \begin{equation}
                ds^2=\Theta(\phi)du^2-2dudr+2\Big(\Xi(\phi)+\frac{u\partial_\phi \Theta(\phi)}{2}\Big)dud\phi+r^2d\phi^2,\ \ \phi\sim\phi+2\pi,
            \end{equation}
            where $u:=t-r$ is the retarded time, $r$ is the radial distance, and $\phi$ is the angle coordinate of the $S^1$ at infinity.
            When $r\rightarrow\infty$, it reaches the null infinity with the coordinates $(\phi,u)$, where the 2d C/G conformal field theory resides.
            In particular, the constant mode solution is expressed as
            \begin{equation}
                ds^2=Mdu^2-2dudr+2Jdud\phi+r^2d\phi^2,\ \ \phi\sim\phi+2\pi,
            \end{equation}
            where the parameters $M$, $J$ are related to the ADM mass $m$ and the angular momentum $j$ of the spacetime,
            \begin{equation}
                m=\frac{M}{8G}, \ \ j=\frac{J}{4G}.
            \end{equation}
            There are two classes depending on the value of $M$:
            \begin{itemize}[label={$\square$}, left=5pt]
                \item \uuline{$M>0$} It is the \textit{flat space cosmological solution} (\textit{FSC}), whose conformal diagram is shown in Fig. \ref{fig:fsc}, with the cosmological horizon at $r_c:=\frac{|J|}{\sqrt{M}}$, which was first considered in string theory context \cite{Cornalba:2003kd}.
                In the $r<r_c$ region, there are closed timelike curves since $\partial_\phi$ is timelike.
                The region $r>r_c$ describes a spacetime in expansion.
                \begin{figure}
                    \centering
                    \begin{tikzpicture}
                        \path[use as bounding box] (-3, -3) rectangle (3, 3);
                        \draw[thick] (0,{2*sqrt(2)}) -- ({-sqrt(2)}, {sqrt(2)}) -- (0,0) -- ({sqrt(2)}, {sqrt(2)}) -- cycle;
                        \draw[thick] (0,{-2*sqrt(2)}) -- ({(-1) * sqrt(2)}, {-sqrt(2)}) -- (0,0) -- ({sqrt(2)}, {-sqrt(2)}) -- cycle;
                        \draw[thick, decorate, decoration={snake, segment length=4pt, amplitude=1pt}] ({-sqrt(2)}, {sqrt(2)}) -- ({-sqrt(2)}, {-sqrt(2)});
                        \draw[thick, decorate, decoration={snake, segment length=4pt, amplitude=1pt}] ({sqrt(2)}, {sqrt(2)}) -- ({sqrt(2)}, {-sqrt(2)});
                        \draw[->, bend left=45] ({-sqrt(2)}, {1.5*sqrt(2)}) to[bend left=50] ({-0.5 * sqrt(2)}, {0.5 * sqrt(2)});
                        \node[left] at ({-sqrt(2)}, {sqrt(2)}) {$i_0$};
                        \node[right] at ({sqrt(2)}, {sqrt(2)}) {$i_0$};
                        \node[left] at ({-sqrt(2)}, {-sqrt(2)}) {$i_0$};
                        \node[right] at ({sqrt(2)}, {-sqrt(2)}) {$i_0$};
                        \node[above] at (0, {2 * sqrt(2)}) {$i^+$};
                        \node[below] at (0, {-2 * sqrt(2)}) {$i^-$};
                        \node[left] at ({-sqrt(2)}, {1.5*sqrt(2)}) {cosmological horizon};
                    \end{tikzpicture}
                    \caption{Flat space cosmological solution (FSC)}
                    \label{fig:fsc}
                \end{figure}
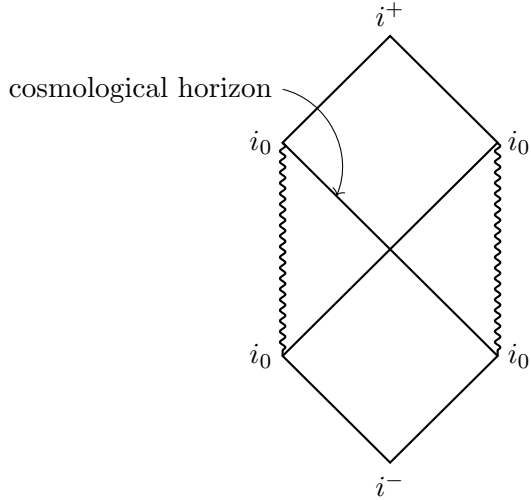
                \item \uuline{$M<0$} It is the spinning particle creating the conical defect and the twist in the time identification.
                They are locally flat but with a delta source.
                The special case is $M=-1$ which is the global Minkowski without conical defect.
                The other cases can be bring back to the Minkowski metric but with a different identification of the coordinates,
                \begin{equation}
                    (u,\phi)\sim(u+2\pi r_0,\phi+2\pi\sqrt{-M})
                \end{equation}
                where $r_0=\frac{J}{\sqrt{-M}}$.
            \end{itemize}
            The holographic entanglement entropy is captured by the swing surface proposal,
            \begin{equation}
                S_A = \frac{A(\gamma_A)}{4G_N},
            \end{equation}
            where $\gamma_A$ is the bench of the swing surface.
            For detailed review, see Appendix \ref{append: swing surface}.
            For the interval $A$ with the length $l_u,l_\phi$ along $u,\phi$ direction,
            in the case $M<0$, the holographic entanglement entropy is
            \begin{equation}
                S_{A}=\frac{1}{4G}\bigg(\sqrt{-M}\Big(l_u-\frac{r_0}{\sqrt{-M}}l_\phi\Big)\cot\frac{\sqrt{-M}l_\phi}{2}+\frac{2r_0}{\sqrt{-M}}\bigg), \label{eq: hee for spining particle}
            \end{equation}
            while in the case $M>0$, the holographic entanglement entropy is
            \begin{equation}
                S_{A}=\frac{1}{4G}\bigg(\sqrt{M}\Big(l_u-\frac{r_c \text{Sign}(J)}{\sqrt{M}}l_\phi\Big)\coth\frac{\sqrt{M}l_\phi}{2}+\frac{2r_c\text{Sign}(J)}{\sqrt{M}}\bigg). \label{eq: hee for fsc}
            \end{equation}
            Now, these results agree with the excited state entanglement entropy expressions given in eqs. \eqref{eq: excited ee1}, \eqref{eq: excited ee2} if we identify
            \begin{equation}
                M<0:\ \   \alpha=\sqrt{-M},\ \ \ \alpha Q=r_0,
            \end{equation}
            \begin{equation}
                M>0:\ \   \alpha'=\sqrt{M},\ \ \ \alpha Q=r_c\text{Sign}(J).
            \end{equation}

            The global Minkowski vacuum has the mass and the angular momentum
            \begin{equation}
                m_{vac}=-\frac{1}{8G}, \ \ j_{vac}=0.
            \end{equation}
            So for $\Xi<\frac{c_M}{24}$, the excited state corresponds to a spinning particle, with the mass $m_p=m-m_{vac}$ and angular momentum $j_p=j-j_{vac}$
            \begin{equation}
                m_p=\Xi,\ \ \ j_p=H.
            \end{equation}
            For $\Xi>\frac{c_M}{24}$, it corresponds to flat cosmological solutions with the mass $m$ and angular momentum $j$
            \begin{equation}
                m=\Xi+\frac{1}{8G},\ \ j=H.
            \end{equation}
            We observed that in the gravitational context, the angular momentum $J$ may assume both positive and negative values.
            This insight motivates the consideration of states and operators with negative conformal weight $\Delta$.
            Notably, the calculations of C/G blocks remain valid even when $H$ is negative.

    \section{Conclusion and discussion}
        In this work, we have advanced the study two-dimensional Carrollian/Galilean Conformal field theories by systematically computing the highest weight conformal blocks in large central charge limit and leveraging them to derive the entanglement entropy of excited states, finding complete agreement with a holographic calculation in three-dimensional asymptotically flat spacetimes in Einstein gravity.

        We successfully calculated the conformal blocks for 2D C/G CFTs in the large central charge limit, focusing on both the \textit{light-type} and the \textit{heavy-light-type} configurations.
        A core implication is that in specific large central charge limits \eqref{eq: large xi}, \eqref{eq: large H}, \eqref{eq: large H Xi}, the full C/G blocks reduce to the global C/G blocks.
        This reduction is achieved by employing a C/G conformal transformation \eqref{eq: sol f}, \eqref{eq: sol g}
        \begin{align}
            f(x) &= 1 - (1 - x)^{\alpha}, \quad g(x) = \alpha Q (1-x)^\alpha \log(1-x),
        \end{align}
        designed to absorb the large charge and weight dependence from the operator product expansions.
        Here, the parameters are defined as,
        \begin{align}
            \alpha &= \sqrt{1 - \frac{24\Xi}{c_M}}, \quad Q = \frac{(\alpha^2 - 1)c_L + 24H}{2\alpha^2 c_M} .
        \end{align}
        Geometrically, this transformation rescales and tilts the cylinder, leading to a new identification $(u,\phi) \sim (u+2\pi\alpha Q,\phi+2\pi\alpha)$.
        Thus, the full C/G block is related to the global block as \eqref{eq: general block to global block 1},
        \begin{equation}
            \lim_{c_M\rightarrow\infty}g(x,y)=(1-w)^{\Delta(1-\frac{1}{\alpha})}\alpha^{2\Delta-\Delta_r}e^{\xi\frac{z(\alpha-1)}{(w-1)\alpha}}g_{\text{global}}(w,z).
        \end{equation}
        From these heavy-light blocks and the assumption of \textit{vacuum block dominance}, we computed the entanglement entropy for states excited by a heavy operator $O_H$.
        For an interval on the plane, the entropy was derived from the $t$-channel vacuum block \eqref{tblock}.
        After mapping to the cylinder, the entanglement entropy of an excited state is identical to the vacuum entanglement entropy on a non-trivially identified cylinder with the rescaled and tilted geometry mentioned above.
        The final result is
        \begin{align}
            S_A = \frac{c_L}{6}\log\left( \frac{2}{\alpha\epsilon} \sin \frac{\alpha l_\phi}{2} \right) + \frac{c_M}{6} \left( Q + \frac{\alpha (l_u - Q l_\phi)}{2} \cot \frac{\alpha l_\phi}{2} \right).
        \end{align}
        Furthermore, when the boost charge exceeds the threshold $\Xi > c_M/24$, the parameter $\alpha$ becomes purely imaginary, $\alpha'=i\alpha$, and the entropy takes a thermal form \eqref{eq: excited ee2}, providing a signature of the Eigenstate Thermalization Hypothesis (ETH) in these theories.

        We further compared our 2d C/G CFT result of the heavy excited state entanglement entropy with the holographic entanglement entropy in 3d flat space cosmological solutions and conical defect solutions given by the swing surface proposal.
        The agreement is perfect through the following dictionary between the C/G CFT and spacetime parameters
        \begin{itemize}
            \item For $M < 0$ solutions (spinning particles/conical defects): $\alpha = \sqrt{-M},\ \alpha Q = r_0$
            \item For $M > 0$ solutions (Flat Space Cosmologies): $\alpha' = \sqrt{M},\ \alpha Q = r_c \text{Sign}(J)$
        \end{itemize}
        This translates to a direct map between the quantum numbers of the excited state and the charges of the spacetime,
        \begin{align}
            m_p &= \Xi, \quad j_p = H \quad \text{(for spinning particles)} \label{eq:particle_dict}\\
            m &= \Xi + \frac{1}{8G}, \quad j = H \quad \text{(for FSCs)}.
        \end{align}

        \subsection*{Discussion and outlook}
            The recent research \cite{Hao:2025btl} suggests that the (light) bulk local excitations are fully reconstructed from the states in the induced representation of the CCFT, while the highest weight representation (HWR) provides limited result.
            There is growing evidence that the induced representation may probe the more fundamental aspects of Flat/CCFT.
            However, in this paper our analysis is heavily reliant on the HWR rather than the induced representation since it is way straightforward to perform explicit calculations of conformal blocks and entanglement entropy in the HWR, leading to the successful holographic match described above.

            One potential physical interpretation is that the HWR characterizes heavy primary states, which induce back reaction effects manifesting as conical defects or FSCs in the bulk, while the induced representation describes light excitations propagating on these backgrounds that do not produce back reaction.
            Another potential interpretation is that despite our calculation of entanglement entropy in the excited state within the HWR, the final result still holds for highly excited states in the induced representation as well.
            The naturally emerging future direction will be to re-examine the calculations of conformal blocks and entanglement entropy using the induced representation, as this may advance a microscopic understanding of flat space holography.

    \acknowledgments
        We are grateful to Liang-Yu Chen,  Luca Ciambelli, Song He, Sabrina Pasterski, Shan-Ming Ruan, Wei Song, Tadashi Takayanagi and Themistocles Zikopoulos for helpful discussions.
        PH is supported by the NSFC special fund for theoretical physics No.\,12447108 and MEXT KAKENHI Grant-in-Aid for Transformative Research Areas (A) through the “Extreme Universe” collaboration: Grant Number 21H05182.
        ST is supported by Division of Graduate Studies Donor Designated Scholarship by Fujitsu Limited and Grant for Overseas Research by the Division of Graduate Studies.

    \appendix

    \section{Swing surface proposal} \label{append: swing surface}
        In this appendix, we detail the holographic entanglement entropy proposal for asymptotically flat three-dimensional spacetimes, so-called the swing surface proposal \cite{Jiang:2017ecm}.
        This formulation provides the flat-space analogue of the RT formula in AdS/CFT.
        We first briefly review the modern formulation of the RT surface in terms of modular flow, and then we present the formal statement and geometric construction of the swing surface from the action of the bulk modular flow.

        \subsection{Modular flow and the Ryu-Takayanagi surface in AdS/CFT}
            The RT proposal $S_A=\frac{A(\gamma_A)}{4G_N}$ posits that the entanglement entropy $S_A$ of a boundary subregion $A$ is proportional to the area of a minimal surface $\gamma_A$ in the bulk that is homologous to $A$ \cite{Ryu:2006bv}.
            A more profound understanding of this prescription arises from the replica trick derivation \cite{Lewkowycz:2013nqa} and the concept of modular flow \cite{Jafferis:2015del}.

            For a given state described by a density matrix $\rho$, the modular Hamiltonian $K_A$ of a subregion $A$ is defined via the reduced density matrix $\rho_A:=\Tr_{A^c}\rho$  as
            \begin{align}
                K_A:=-\log\rho_A.
            \end{align}
            The unitary evolution $U(s) = e^{-isK_A}$ is called the modular flow.
            In the context of AdS/CFT, for a state dual to a classical geometry, the boundary modular Hamiltonian $K_A$ has a corresponding bulk operator \cite{Jafferis:2015del}
            \begin{equation}
                K_A = \frac{\mathcal{A}(\gamma_A)}{4G_N} + K_{\text{bulk}} + \dots,
            \end{equation}
            where $\mathcal{A}(\gamma_A)$ is the area operator of the RT surface $\gamma_A$, and $K_{\text{bulk}}$ is the bulk modular Hamiltonian associated with the entanglement wedge $\mathcal{E}_A$.

            The bulk modular flow generated by $K_{\text{bulk}}$ manifests as a geometric flow within the entanglement wedge $\mathcal{E}_A$.
            The RT surface $\gamma_A$ is precisely the region left invariant by the bulk modular flow.
            In the standard AdS/CFT setup, the boundary $\partial A$ is also a fixed point set of the boundary modular flow, and the minimal surface $\gamma_A$ anchors directly on it, such that $\partial\gamma_A = \partial A$.
            This correspondence is altered in the case of asymptotically flat spacetimes.

        \subsection{Swing surface for holographic entanglement in flat space}
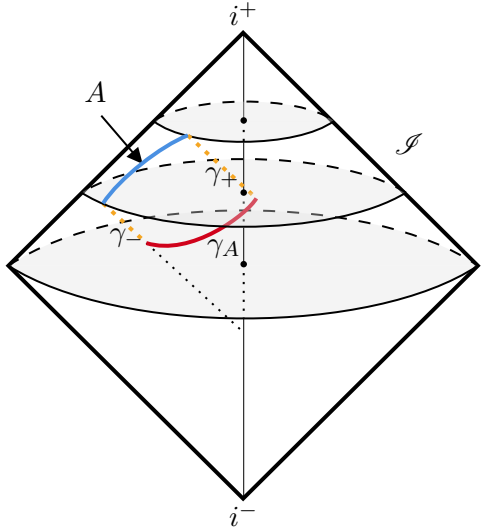
\begin{figure}[t]
                \centering
                \tikzset{every picture/.style={line width=0.75pt}} 
                \begin{tikzpicture}[x=0.75pt,y=0.75pt,yscale=-0.9,xscale=0.9]
                    \draw  [line width=1.5]  (349.68,10.68) -- (479.86,140.86) -- (349.68,271.03) -- (219.5,140.86) -- cycle ;
                    \draw[fill={rgb, 255:red, 225; green, 225; blue, 225 },fill opacity=0.35]     (219.5,140.86) .. controls (281,180.15) and (421,180.15) .. (479.86,140.86) ;
                    \draw  [dash pattern={on 4.5pt off 4.5pt},fill={rgb, 255:red, 225; green, 225; blue, 225 },fill opacity=0.35]  (219.5,140.86) .. controls (281,99.15) and (421,100.15) .. (479.86,140.86) ;
                    \draw[line width=0.3]    (350,140) -- (350,119.57) ;
                    \draw [line width=0.3]    (350,10.68) -- (350,59.67) ;
                    \draw[line width=0.3]    (350,170) -- (350,271.03) ;
                    \draw[dash pattern={on 0.84pt off 2.51pt}]  (350,140) -- (350,170) ;
                    \draw[fill={rgb, 255:red, 225; green, 225; blue, 225 },fill opacity=0.35]    (300,60) .. controls (320,76) and (380,75) .. (400,60) ;
                    \draw[dash pattern={on 4.5pt off 4.5pt}, fill={rgb, 255:red, 225; green, 225; blue, 225 },fill opacity=0.35]  (300,60) .. controls (320,46) and (380,45) .. (400,60) ;
                    \draw[dash pattern={on 4.5pt off 4.5pt}, fill={rgb, 255:red, 225; green, 225; blue, 225 },fill opacity=0.35]  (440,100) .. controls (400,75) and (301,75) .. (260,100) ;
                    \draw [line width=0.3]    (350,72) -- (350,100) ;
                    \draw  [dash pattern={on 0.84pt off 2.51pt}]  (350,100) -- (350,119.57) ;
                    \draw  [dash pattern={on 0.84pt off 2.51pt}]  (350,60) -- (350,72) ;
                    \draw [color={rgb, 255:red, 245; green, 166; blue, 35 }  ,draw opacity=1 ][line width=1.5]  [dash pattern={on 1.69pt off 2.76pt}]  (272,106) -- (296,128.2) ;
                    \draw [color={rgb, 255:red, 245; green, 166; blue, 35 }  ,draw opacity=1 ][line width=1.5]  [dash pattern={on 1.69pt off 2.76pt}]  (319,68) -- (357,103.2) ;
                    \draw  [dash pattern={on 0.84pt off 2.51pt}]  (296,128.2) -- (350,178.2) ;
                    \draw[color={rgb, 255:red, 208; green, 2; blue, 27 }  ,draw opacity=1 ][line width=1.5]    (296,128.2) .. controls (321,136) and (352,112) .. (357,103.2) ;
                    \draw[fill={rgb, 255:red, 225; green, 225; blue, 225 },fill opacity=0.35] (440,100) .. controls (400,125) and (300,126) .. (260,100) ;
                    \draw    (271,57) -- (292.05,81.72) ;
                    \draw [shift={(294,84)}, rotate = 229.57] [fill={rgb, 255:red, 0; green, 0; blue, 0 }  ][line width=0.08]  [draw opacity=0] (8.93,-4.29) -- (0,0) -- (8.93,4.29) -- cycle;
                    \draw [color={rgb, 255:red, 74; green, 144; blue, 226 }  ,draw opacity=1 ][line width=1.5]    (272,106) .. controls (281,92) and (304,74) .. (319,68);
                    \fill[black] (350,140) circle[radius=1.8];
                    \fill[black] (350,59.67) circle[radius=1.8];
                    \fill[black] (350,100) circle[radius=1.8];
                    \draw (432,65.4) node [anchor=north west][inner sep=0.75pt]    {$\mathscr{I}$};
                    \draw (350,0) node [inner sep=0.75pt]    {$i^+$};
                    \draw (350,280) node [inner sep=0.75pt]    {$i^-$};
                    \draw (327,84.4) node [anchor=north west][inner sep=0.75pt]    {$\gamma _{+}$};
                    \draw (274,118.4) node [anchor=north west][inner sep=0.75pt]    {$\gamma _{-}$};
                    \draw (328,124.4) node [anchor=north west][inner sep=0.75pt]    {$\gamma _{A}$};
                    \draw (260,36.4) node [anchor=north west][inner sep=0.75pt]    {$A$};
                \end{tikzpicture}
                \caption{The geometric construction of the swing surface.
                    The entanglement entropy of the boundary interval $A = [p_1, p_2]$ at null infinity $\mathscr{I}$ is given by the length of the bulk spacelike geodesic $\gamma$ (the swing), which connects to the endpoints of $A$ via two null geodesics $\gamma_\pm$ (dashed orange).}
                \label{fig:swing_surface}
            \end{figure}

            In flat space holography, the boundary is null infinity $\mathscr{I}$.
            A spacelike bulk surface cannot anchor on a null boundary without violating its spacelike character at the intersection.
            The swing surface proposal \cite{Jiang:2017ecm,Apolo:2020bld} provides the modification through a composite bulk surface, depicted in Figure \ref{fig:swing_surface}.

            The construction of the swing surface involves three components.
            \begin{enumerate}
                \item A spacelike geodesic $\gamma$ located deep in the bulk.
                This component is referred to as the \textit{swing} or \textit{bench}.
                \item Two null geodesics $\gamma_+$ and $\gamma_-$ that extend from the endpoints of the bulk geodesic $\gamma$ to the endpoints of the boundary interval $A$ on $\mathscr{I}$.
                These are the \textit{ropes}.
            \end{enumerate}
            The crucial distinction from the RT proposal is that the bulk spacelike surface $\gamma$ does not anchor directly on the boundary $\partial A$.
            Instead, it is connected to $\partial A$ via the null ropes $\gamma_\pm$.

            The physical origin of this detached structure lies in the distinct actions of the boundary and bulk modular flows.
            While the endpoints of the boundary interval $\partial A$ are fixed points of the \textit{boundary} modular flow, they are not fixed points of the \textit{bulk} modular flow \cite{Apolo:2020qjm}.
            The swing surface construction provides a precise geometric interpretation for this mismatch.

            \begin{tcb}[Swing surface proposal]
                The holographic entanglement entropy for a boundary interval $A$ is given by the area (length in 3d) of the spacelike geodesic $\gamma_A$ (the swing).
                \begin{equation}
                    S_A = \frac{A(\gamma_A)}{4G_N}.
                \end{equation}
                The lengths of the ropes $\gamma_\pm$ do not contribute to the entropy since they are null.
            \end{tcb}

            \paragraph{The swing as the fixed-point set.}
                The spacelike geodesic $\gamma$ is the set of fixed points of the bulk extended modular flow, generated by the bulk modular operator $K_{\text{bulk}}$.
                It is the direct flat-space analogue of the RT surface.
                Just as the RT surface is the minimal area surface, the swing $\gamma$ is the extremal length spacelike geodesic whose endpoints are determined by the termination of the ropes.

            \paragraph{The ropes as integral flows.}
                The null geodesics $\gamma_\pm$ are the integral curves of the boundary endpoints $\partial A = \{p_1, p_2\}$ under the action of the bulk modular flow.
                They trace the path of the boundary endpoints as they are evolved by the bulk's geometric flow.
                This evolution is non-trivial because, as mentioned, the points in $\partial A$ are not fixed by the bulk flow.

            \paragraph{Junction points.}
                The ropes $\gamma_\pm$ originate at the boundary points $p_1, p_2 \in \partial A$ and terminate at the bulk points $q_1, q_2 \in \partial\gamma$.
                These junction points $\{q_1, q_2\}$ are precisely where the non-trivial orbits of the bulk modular flow, initiated at the boundary, intersect the set of fixed points of that same flow.
                The swing $\gamma$ is then the minimal length spacelike geodesic connecting these two termination points.
                This construction geometrically decomposes the action of the modular operator.
                The ropes represent the dynamical ``flow" part of the evolution, while the swing represents the stationary ``fixed point" or ``horizon" part, whose length gives the entropy.

                This entire structure can be derived more formally from the Rindler method.
                By finding a coordinate transformation that maps the entanglement calculation to a thermal one, the entanglement wedge is mapped to a bulk Rindler wedge possessing a Rindler horizon.
                The length of this Rindler horizon, which is by definition a fixed-point set of the associated boost symmetry, exactly corresponds to the length of the swing $\gamma$.
                This confirms that the swing surface is the appropriate ``entanglement horizon" for a subregion on a null boundary.

    \bibliographystyle{JHEP}
    \bibliography{main}

\end{document}